\documentclass[english,onecolumn]{elsart3p}
\usepackage{amssymb}
\usepackage{amsmath}
\usepackage{babel}
\usepackage{graphicx}

\newcommand{\vect}[1]{\boldsymbol{\mathrm{#1}}}

\newcommand{\abs}[1]{\vert#1\vert}

\begin{document}

\begin{frontmatter}
	\title{Rough-Surface Shadowing of Self-Affine Random Rough Surfaces}

  \author{Hannu Parviainen}, \ead{hannu@astro.helsinki.fi}
	\author{Karri Muinonen} \ead{muinonen@cc.helsinki.fi}
  \address{Observatory, B.O.\ Box 14, FI-00014 University of Helsinki, Finland}

  \begin{abstract}
  Light scattering from self-affine homogeneous isotropic random rough surfaces is studied using the
ray-optics approximation. Numerical methods are developed to accelerate the first-order scattering
simulations from surfaces represented as single-connected single-valued random fields, and to store
the results of the simulations into a numerical reflectance model. Horizon mapping and marching
methods are developed to accelerate the simulation. Emphasis is given to the geometric shadowing and
masking effects as a function of surface roughness, especially, to the azimuthal rough-surface
shadowing effect. 

  \end{abstract}

\end{frontmatter}

\section{Introduction}
\label{sec:intro}
Scattering of light from rough surfaces has been a topic of extensive study for several
decades. Both theoretical and experimental research is carried out in many areas of physics.  
The amount of empirical research is vast, including both observational data---from 
Earth-based to space-probe observations---as well as strictly controlled laboratory measurements, 
and the increase of computational power in last decades has allowed application of advanced
numerical methods. 

Despite the advancement of the analytic rough-surface light scattering models, they are
limited by necessity to simplified surface-roughness structures. Exact solutions to the
rough-surface scattering have been derived for simple periodic structures, but realistic surfaces
occurring in nature cannot be well represented with such.
Natural surfaces show roughness at all scales---from the large-scale variations describing the
general shape of the object down to the molecular, and ultimately atomic, structure of the medium.
These roughness variations over different size scales can often be considered to be of self-affine
fractal nature: the resolved roughness is a function of the scale of the observation.
More realistic scattering models have been devised for several statistical random rough surface
models, but fractal surfaces are still beyond analytic approach. Currently numerical simulation
is the most valid method to test the applicability of the analytic scattering models, and
the only method when it comes to the understanding of scattering by complex self-affine rough
surfaces.

The theory of light scattering from planetary regoliths has been studied in depth by Lumme, Bowell
and Irvine \cite{Lumme81a,Lumme82}, who introduced a scattering model for a rough cratered surface
and tested its validity against observational data. Hapke introduced his own analytic
reflectance model derived from the equation of radiative transfer \cite{Hapke81a}, with a correction
for macroscopic surface roughness \cite{Hapke84}. Later Lumme, Peltoniemi and Irvine \cite{Lumme90}
introduced a model for diffuse reflection from stochastically bounded semi-infinite medium based on
Gaussian statistics, which was used to derive an average single-particle phase function
for the lunar regolith \cite{Lumme89}.
Numerical studies have been carried out by Peltoniemi \cite{Peltoniemi93}, who studied light
scattering from closely-packed particulate medium. Shkuratov et al. have used numerical simulations
together with photometric and polarimetric laboratory measurements of samples simulating the
structure of planetary regoliths \cite{Shkuratov02,Shkuratov04}. Shepard and
Campbell \cite{Shepard98,Shepard1999,Shepard2001} have studied geometric shadowing from fractal
surfaces. Their study was focused on the behavior of the shadowing function when the surface is
viewed from nadir, and showed that many of the currently used analytical shadowing functions can be
well fitted to results obtained from self-affine fractional Brownian motion (fBm) surfaces. The
roughness of natural surfaces has been shown to  often follow fractal self-affine statistic,
spanning all the observable scales. Thus, the fBm-model has been shown to allow for more realistic
modeling of natural rough surfaces than the more frequently used models with a single major scale
for roughness features.

The rough-surface shadowing and masking effects on light scattering are studied in this paper as a
function of the angle of incidence $\theta_i$ and surface roughness model specific parameters $(P_1,
P_2)$ (described later in more detail) within the limits of the first-order geometric-optics
approximation. All of the interactions of a single ray are considered to take place inside a single
surface element, and the size of the surface curvature is considered to be substantially larger than
the wavelength of the radiation. Special emphasis is given
to the azimuthal shadowing and masking effect, caused by the self-shadowing of the surface. A
comparison is made between two different surface roughness models: Gaussian correlation and
generalized fractional Brownian motion (fBm). The Gaussian correlation model is good for its
simplicity, but is not well fit to mimic natural rough surfaces, for which the self-affine fractal
fBm-surfaces are more suited. The simulation is calculated over the full hemisphere $(\theta_e,
\phi_e)$ discretized using recursive spherical quadtree model for each set of free simulation
parameters $(\theta_i, P_1, P_2)$. 
The results over the hemisphere are saved into a numerical scattering model which allows
interpolation inside the five-dimensional parameter space $(P_1, P_2, \theta_i, \theta_e, \phi_e)$
mixing spherical and Cartesian bases. The model can be used in varying inversion problems, such as
statistical photoclinometry \cite{Muinonen1990} and asteroid studies, to estimate the surface
roughness properties.

\section{Theory}
\label{sec:theory}
\subsection{Geometric Rough-Surface Shadowing and Masking Effects}\label{ss:geometric_effects}

In the regime of geometric optics, surface roughness has a notable effect on the 
radiation scattered from the surface. The surface features can shadow the surface
from incident radiation, and mask the reflection from the irradiated parts. 
For a homogeneous isotropic surface with certain roughness statistics, the geometric 
self-shadowing and self-masking can be combined into a shadowing/masking 
function $S(\theta_i, \phi_i, \theta_e, \phi_e)$, which in the 
single-scattering approximation is independent of the underlying reflectance model. 
The shadowing/masking function gives the probability for a point visible to the observer
to be illuminated as a function of the incidence angles $(\theta_i, \phi_i)$
and the emergence angles $(\theta_e$, $\phi_e)$  \cite{Penttila05}. 

Exact formulation of the combined shadowing and masking function is currently
possible only for a few simplified surface roughness models, and a numerical
simulation is necessary when more realistic surfaces are of interest. Roughness
models with self-affine properties, especially suited for mimicking many complex
natural surfaces, are of special interest.

While a spectrum of advanced reflectance models have been developed with geometric shadowing as a 
function of $\theta_i$ and $\theta_e$ calculated by analytical means for several types of surface
roughness, the $\phi_e$ dependence has generally been left out from the calculations because of the 
major increase in mathematical complexity. Neither do these reflectance models take the fractal
nature of the surface roughness into account. While analytically
difficult to model, surfaces with fractal properties can be simulated numerically, and the $\phi_e$
dependence can be included in the simulations. Numerical simulations allow the study of the
azimuthal shadowing effect as a function of surface roughness, and the generation of a numerical
rough-surface scattering model which can be evaluated as a function of viewing geometry, and the
surface-roughness statistics.

\subsection{Numerical Rough-Surface Reflectance Model}\label{ss:reflectance}

When the light scattering properties of a single surface element are considered to follow
from a theoretical scattering model, a numerical rough-surface-corrected version of the model 
can be obtained from the simulations. Now, in addition to the shadowing and masking effects,
the distribution of normals of the irradiated surface elements visible to a certain direction
will alter the total observed reflectance of the surface.

Bidirectional reflectance-distribution function (BRDF) \cite{Nicodemus77,Lester79,Parviainen06} 
$f_r$ is defined as the ratio between the reflected radiance $L_r$ and total irradiance $E$ of 
the surface element as
\begin{equation}
	f_r(\theta_i, \phi_i, \theta_r, \phi_r) =  \frac{L_r(\theta_i, \phi_i, \theta_r,
\phi_r)}{E}.
\end{equation}
\noindent Considering unidirectional incident radiation $(E = \mu_0 F)$ and isotropic surfaces, for
which we
can use symmetry relations to remove the explicit $\phi_i$ and $\phi_r$ dependence with their
difference $\Delta \phi$, BRDF can be expressed as
\begin{equation}
	f_r(\mu_0, \mu, \Delta \phi) =  \frac{L_r(\mu_0, \mu, \Delta \phi)}{\mu_0 F},
\end{equation}
\noindent where $\mu_0 = \cos \theta_i$ and $\mu = \cos \theta_r$. Two basic reflectance functions 
are considered in current paper: the perfect 
diffuse Lambertian reflectance function $f_{r,L}$, and a more realistic Lommel-Seeliger
reflectance function $f_{r,LS}$. These are respectively
\begin{align}
	f_{r,L}  &= \frac{\lambda_0}{\pi} \mu_0 F,\\
	f_{r,LS} &= P(\alpha) \frac{\tilde{\omega}}{4 \pi} \frac{\mu_0}{\mu + \mu_0} F,
\end{align}
\noindent where $P(\alpha)$ is the phase function, $\tilde{\omega}$ the single scattering
albedo, and $\lambda_0$ the Lambertian albedo. Since $\lambda_0$ and $\tilde{\omega}$ are nothing
but multiplicative factors, when considering the first-order scattering, they are set to
unity in the study. The phase function $P(\alpha)$ is also set to constant $P(\alpha) = 1$ to
constrain the number of free parameters of the scattering model.

The major problem with a numerical reflectance model is how one can store and
interpolate a five-dimensional function ($[H,l], \sigma, \theta_i, \theta_r, \Delta \phi$),
where $H$, $l$ and $\sigma$ are surface roughness parameters, of both Cartesian and 
spherical spaces with sufficient angular resolution. Several methods have been
implemented to overcome this problem in previous studies, including spherical and
hemispherical harmonics \cite{Sillion91,Westin92a,Westin92b,Gautron04}, 
Zernike polynomials \cite{Koenderink96}, and spherical wavelets \cite{Schroder95a}.
Hemispherical harmonics \cite{Gautron04} was chosen for their computational efficiency
and close relation to the well-known spherical harmonics.

\subsection{Surfaces}\label{ss:surfaces}
Surfaces are represented as two-dimensional isotropic homogeneous random 
fields \cite{Adler81,Preston76,Ibragimov70} specified by the autocorrelation 
function $C(r)$ \cite{Fenton90a} or, equivalently, spectral density function $S(f)$
\cite{Peitgen88}.
The distribution of heigths follows Gaussian statistics, and is defined by the
standard deviation $\sigma$. The field realizations are periodic in $x$ and $y$,
with length of the period $L$.

Generation of a surface realization is based on the spectral synthesis method \cite{Fenton90a,Parviainen06,Dieker02}.
The surface statistics determine the power spectrum of the surface features in the 
frequency domain, and a surface realization is generated by transforming a 
randomized realization of the power spectrum to the spatial domain using fast Fourier
transform (FFT). 

\begin{figure}[h!]
	\centering
	\includegraphics[width=\textwidth]{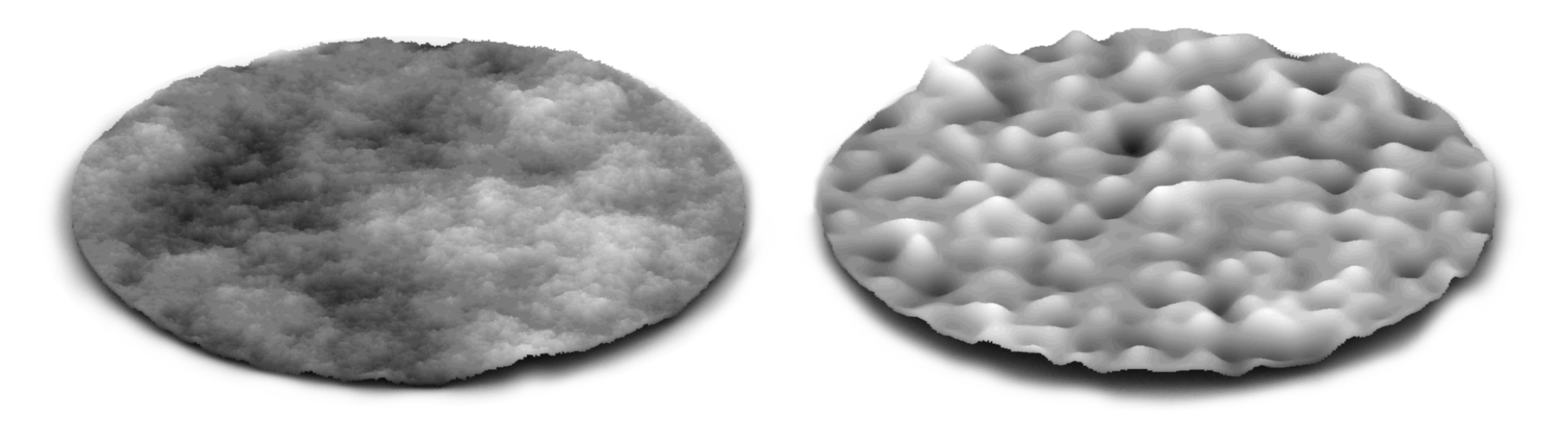}
	\caption{An fBm-surface realization with $H = 0.5$ and a surface realization with Gaussian
	correlation with $\frac{l}{L} = 0.025$. The standard deviation of heights $\frac{\sigma}{L} =
	0.01$ is equal for both surfaces.}
	\label{fig:surfaces}
\end{figure}

Two different roughness types were chosen for the study: Gaussian correlation
and general fractional Brownian motion models. Surfaces with Gaussian correlation 
between drops show roughness features of a certain scale, determined by the correlation
length $l$, 
while the fBm-surfaces are of self-affine fractal nature showing 
roughness features of all scales, determined by the Hurst exponent $H$. 
The difference between the two models is illustrated in Fig. \ref{fig:surfaces}, where the fBm
surface and Gaussian surface realization have identical $\frac{\sigma}{L}$.

The radially symmetric spectral density functions for the Gaussian and fBm roughness models are
\begin{align}
	S_{Gauss}(\vect{k}) &= \frac{l}{2 \sqrt{\pi}} \exp(-\frac{\abs{\vect{k}}^2 l^2}{4}),\\
	S_{fBm}(\vect{k})   &= \abs{\vect{k}}^{-(H+1)},
\end{align}
\noindent where $\abs{\vect{k}}$ is the wavenumber. The correlation length is the distance where
the correlation between drops down to $e^{-1}$ and the Hurst exponent is related to the fractal 
dimension $D$ of the surface as $H = 3 - D$.

\section{Numerical Methods}
\label{sec:methods}

\subsection{Overview}

The simulation is carried out in the ray-optics regime. The surface curvature radii in the
simulation are considered to be much larger than the wavelength of light (tangent-plane
approximation), and all the interactions of a single ray of light can be considered to take place
inside a single surface element (first-order scattering approximation). For numerical calculations,
both
the surface and the integrating hemisphere are tessellated into discrete polygonal elements. The
scattering model space is discretized for the $\theta_i$ angle of the incident radiation, and for
the two surface roughness parameters $([H,l], \sigma)$.

The simulation consists of selecting $n_{p}$ sample points $p_j$ from a surface realization $s_i$,
computing the highest unshadowed value $\theta_i$ for $p_j$, computing the solid area elements
of the integrating hemisphere visible from $p_j$, and adding the contribution $F_{pj}(\theta_i,
\phi_i, \theta_e, \phi_e)$ to each of the visible solid area element for each discrete value of
$\theta_i$. The simulation is repeated for $n_{s}$ surface realizations, and the outcome is the 
ensemble average of the radiance from surface points visible to a direction of each of the
solid area elements as a function of $\theta_i$. If there is no BRDF assigned to the elements of the
surface, the output is the geometric shadowing and masking function. Finally, the output is expanded
into hemispherical harmonic coefficients for each $\theta_i$, and the coefficients
are saved to represent the model for the chosen set of parameters $H$ and $\sigma$.

While the entire simulation could be done using basic ray-tracing techniques, the use of
specialized optimization methods can increase the computation speed dramatically. The decreased
computation time allows for more accurate simulations, and a reliable study of different effects as
a function of the surface roughness parameters.
The surface sample points $p_j$ are selected from the surface plane using stratified Monte Carlo
sampling \cite{Pharr04}, which offers a good trade-off between speed and the quality of sampling.
A horizon map generated for the surface is used to find the highest unshadowed value for the angle
of the incident radiation $(\theta_i)$ for each sample point $p_j$. The integrating hemisphere is
represented as a recursive hierarchical geometry with discrete elements of solid area, and a
horizon-marching method is implemented for it to accelerate the ray-tracing process.

\subsection{Surfaces}

The surface is triangulated for fast ray-triangle intersection tests into a uniform triangular grid
which allows for grid tracing \cite{Musgrave89} method to be used with run-based
ray-traversal \cite{Stephenson98,Stephenson00,Stephenson01} for optimal speed. Horizon mapping is
used extensively in the simulation, and several optimized methods to calculate the horizon height to
a certain direction are implemented for different situations.

Whether a point $P(x,y)$ on a surface is shadowed from unidirectional radiation $L(\theta_i,
\phi_i)$ can be tested in several ways. The basic ray-tracing method is simply to test if a ray
from $P$ can escape the surface to the direction of $L$ without intersecting any other
surface element. Nevertheless, ray tracing the intersection test for all $(\theta_i, \phi_i)$ is
computationally heavy and, in the end, not necessary. More efficient method is to calculate
for $P$ the height of the horizon to the direction $\phi_i$, i.e., the highest unshadowed value
$\theta_i$. When the height of the horizon $\theta_H(\phi)$ is known, we are able to tell if the
point is in shadow for arbitrary $\theta_i$. 

Two methods for the calculation of $\theta_H(\phi)$ were implemented. For Gaussian surfaces
with continuous curvature, the horizon height in $\phi_i = 0$ direction can be calculated from the
height-data of the surface vertices along the $x$-axis, and no ray-tracing methods need to be used.
The height of the horizon for a point inside a surface element can be interpolated from the
$\theta_H$ calculated for the vertices of the surface element. For discontinuous fBm-surfaces this
method is not reliable, and ray-tracing is used to search for accurate $\theta_H$.

\subsection{The Integrating Hemisphere}
\subsubsection*{Hemisphere Representation}

\begin{figure}[h!]
	\centering
	\includegraphics[width=0.9\textwidth]{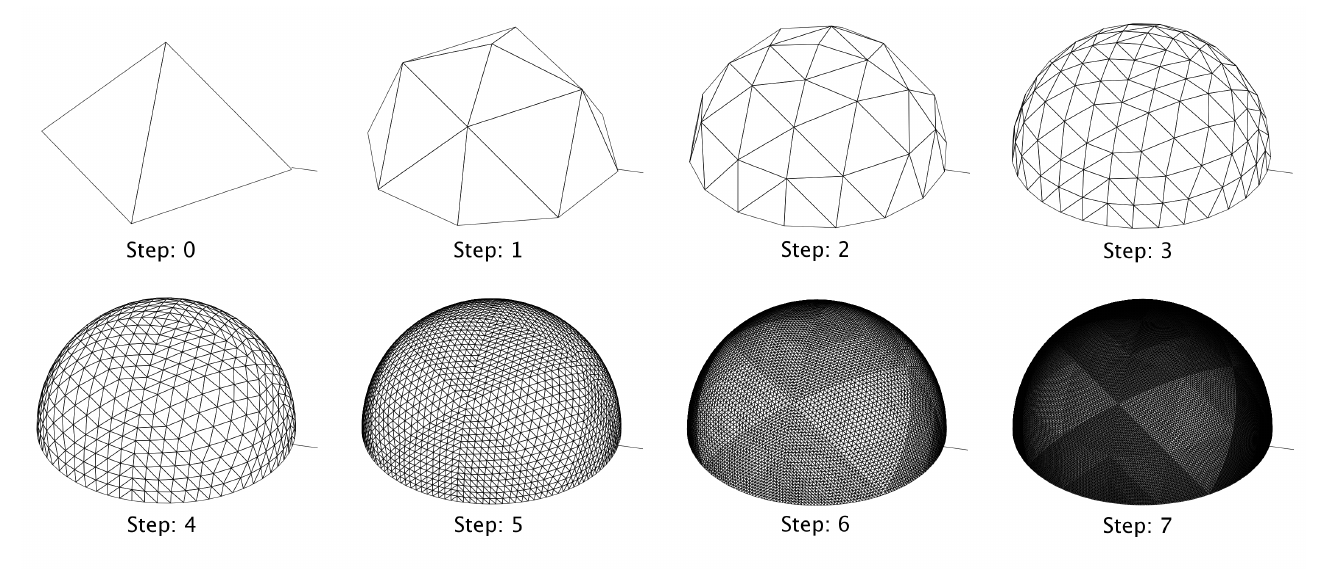}
	\caption{The integrating hemisphere shown with increasing number of subdivisions.
	During each subdivision, all of the facets (nodes) of the hemisphere are subdivided into
	four subfacets, and the positions of the new vertices are normalized. This leads to an uneven
	distribution of solid angles,
	but is nevertheless better than sampling along constant angles of $\theta_e$ and $\phi_e$}
	\label{fig:hemisphere_tesselation}
\end{figure}

The integrating hemisphere stores the scattering data during the simulation into a quadtree-based
data-structure \cite{Barrett95}. The tessellation of the hemisphere is carried out  with methods
based
on triangular hierarchical data structures as discussed by Goodchild \cite{Goodchild91},
Dutton \cite{Dutton96}, Sahr \cite{Sahr03}, and Kunszt \cite{Kunszt01}. The upper hemisphere of an
octahedron was chosen to form the root of the quadtree, and the subnodes are obtained by subdivision
of the geometry (see Fig. \ref{fig:hemisphere_tesselation}) using Class I subdivision
scheme \cite{Sahr03}. The main difference between the current implementation and the cited previous
implementations is the use of a half-edge structure to represent the geometry
with \cite{Parviainen06}. This adds some extra memory overhead to the system, but makes the
recursive
subdivision of the underlying geometry fast, and simplifies the task of finding neighbors for
different parts of the structure.

Two different hemisphere sampling strategies are used. For relatively smooth surfaces, a 
horizon marching method \cite{Parviainen06} is utilized to accelerate the search of the 
elements of the hemisphere, where the radiation may scatter directly from the current surface 
sample point. For extremely rough surfaces, where the horizon marching method is not applicable,
each of the hemisphere elements is sampled separately.

\subsubsection*{Horizon Marching}

Since the surface height is represented by a single-valued single-connected function of two
dimensions $(x,y)$, the height of the horizon $\theta_h$ for a point on the surface is a
single-valued continuous function of the direction $\phi$, $H(\phi) \simeq \theta_h$.
The hemisphere from a point on the surface can be factorized into two parts, divided by the
horizontal height function. The first part consisting of the directions culled by the surface, and
the other of the directions where the radiation can escape directly. If $H = H(\phi)$ can be
calculated for a surface sample point, we can immediately evaluate the scattering function to the
upper hemisphere, without the need to test each single element for occlusion. For a random rough
surface, $H(\phi)$ for an arbitrary point on the surface cannot be calculated analytically, and
approximate methods have to be used.

The approximation for a single irradiated surface sample point is started by searching the height of
the horizon at $\phi = 0$ in the space of the tessellated hemisphere geometry. After the value for
$H(\phi = 0)$ is found, a marching phase starts. The octahedral geometry combined with the half-edge
data structure allows one to travel in the hemisphere geometry in an efficient manner.
A vertex in the hemisphere geometry is in general connected to six other vertices, the pole,
corners, and edges of the octahedron working as special cases. Three of these are in the direction
of positive $\phi$, and three of negative $\phi$. From the first horizon vertex $v_{h, \phi=0}$, we
find the lowest unoccluded neighbor vertex in the direction of positive $\phi$ by testing the three
possible traveling directions. The process is repeated from the newfound horizon vertex, and the
approximation to $H(\phi)$ is found after marching around the hemisphere. Finally, a scanline-type
method is used to select the solid area elements over the horizon.

\begin{figure}[h!]
	\centering
	\includegraphics[width=1.0\textwidth]{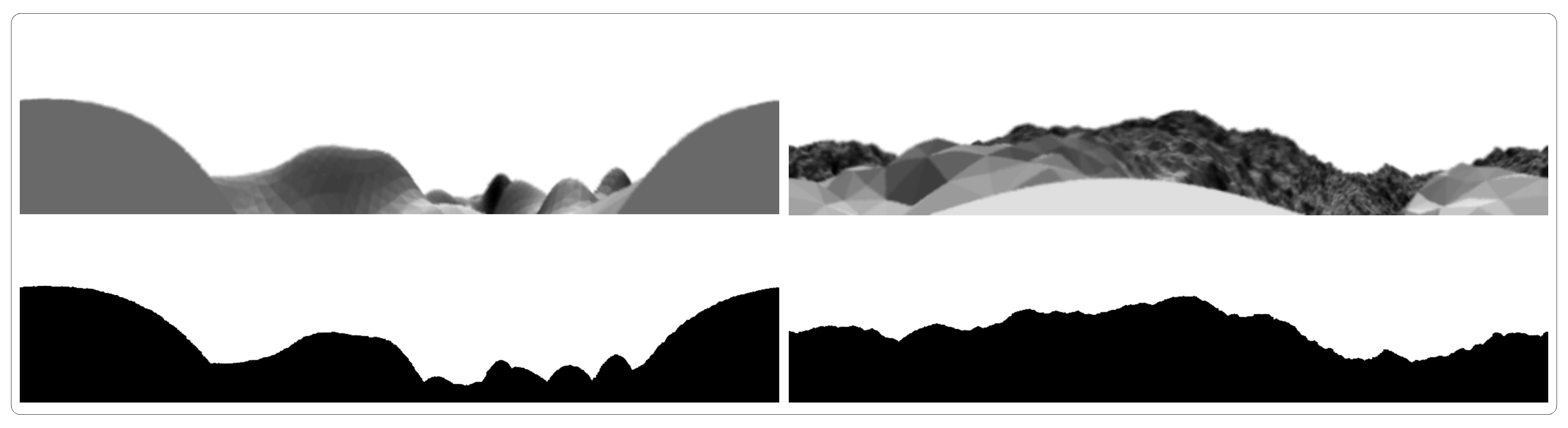}
	\caption{Projections of a rough surface from a single surface sample point to the integrating
hemisphere.
	On the left we show a surface following Gaussian correlation and, on the right, a surface with fBm
	statistics. The upper subplots show a projection using a full sampling method, and a
	Lommel-Seeliger scattering law applied to the surface elements. The lower subplots show the
	horizon of the same points traced using the horizon marching method.}
	\label{fig:horizons}
\end{figure}

While more complex to implement than the Monte-Carlo or full sampling of the hemisphere, the
computation speed is superior to both. The resulting accuracy is comparable to the full sampling,
but the number of trace calls needed is around $\frac{1}{20}$ of the full sampling method. The
current implementation of the horizon marching method is 12 times faster than the full sampling
method implemented. Only when the vertical variation of the horizon line becomes extreme compared
to the hemisphere tessellation resolution, i.e. for very rough surfaces, the horizon marching method
fails. This can be corrected by adding more possible directions for the method to choose from, thus
allowing more complex shapes to be traced, or by changing to the full sampling method.

\section{Results and Discussion}
\label{sec:results}

We ran the simulations for three distinct BRDF cases. First, we consider geometric shadowing and
masking effects independent of any scattering model, next we study the first-order rough-surface
scattering effects arising for surfaces with Lambertian scattering elements and, finally, a more
realistic case with surfaces consisting of Lommel-Seeliger scattering elements. The Lommel-Seeliger
scattering model is of interest on a physical basis, while the Lambertian scattering model is good
for comparison with the study by Shkuratov et al. \cite{Shkuratov04}. Two different
surface-roughness statistics were used: the Gaussian correlation model and the fBm model. Both
models were sampled in the parameter space, and the results for three different
$H$ and $l$ are presented, each with three values for $\frac{\sigma}{L}$. The advantage of two
fundamentally different surface roughness models is clear: Gaussian surfaces allow us to study the
effects arising from a single well-defined scale of roughness features, while fBm surfaces
show the effects for surfaces with roughness of all scales, down to the tessellation size of the
surface.

The surfaces were generated as $1024 \times 1024$ point regular grids with period $L$ of $100.0$
units. This leads to the minimum facet edge scale of $0.98$ units. The values used for $\sigma$ were
0.5, 1.5, 2.5, leading to $\frac{\sigma}{L} = 0.005, 0.015, 0.025$, where $\frac{\sigma}{L} =
0.025$ corresponds to an extremely rough surface. For Gaussian surfaces, the correlation lengths
$l$ chosen for the study are 1.0, 5.0, 10.0, while for fBm-surfaces we used
Hurst-exponents H of $0.3, 0.5, 0.7$. The simulations were ran with 200 surface realizations for
each set of parameters $([H,l], \sigma)$, and 100 sample points for each realization. Extremely
rough surfaces ($H=0.3$ or $\frac{\sigma}{L} = 0.025$) were simulated using 400 surface realizations
and $1600 \times 1600$ point surface resolution to ensure good statistics over the whole
integrating hemisphere.

The integrating hemisphere was set to subdivision level 6 (see Fig.
\ref{fig:hemisphere_tesselation}), which results to a mean solid angle per hemisphere facet of
$1.4$ square arcsecond. Along the constant lines $\phi_e = \frac{n\pi}{2}$, we get
$\frac{\pi}{128} \approx 0.025$ radians for the $\theta_e$ resolution. Since the major interest of
the study is the azimuthal shadowing effect near the back-scattering direction, the results shown 
are restricted to $\theta_e = [0, \frac{\pi}{2}]$ and $\phi_e = [0, \frac{\pi}{2}]$.

The simulation accuracy was tested by comparing the simulation results against the results obtained
from $2048 \times 2048$ point surfaces using subdivision level 7 for the integrating hemisphere.
The accuracy did not increase significantly for the test cases used.

\subsection{Shadowing and Masking}

First, we consider pure shadowing and masking effects arising from the surface roughness.
The shadowing function $S(\theta_i, \theta_e, \Delta \phi)$ gives us the probability that a surface
point is both visible and not shadowed. Geometrically this is equivalent to the ratio of the
irradiated surface area projected to a plane with normal $(\theta_e, \phi_e)$, to the total 
surface area projected to the same plane.

\begin{figure}
	\centering
	\includegraphics[width=0.85\textwidth]{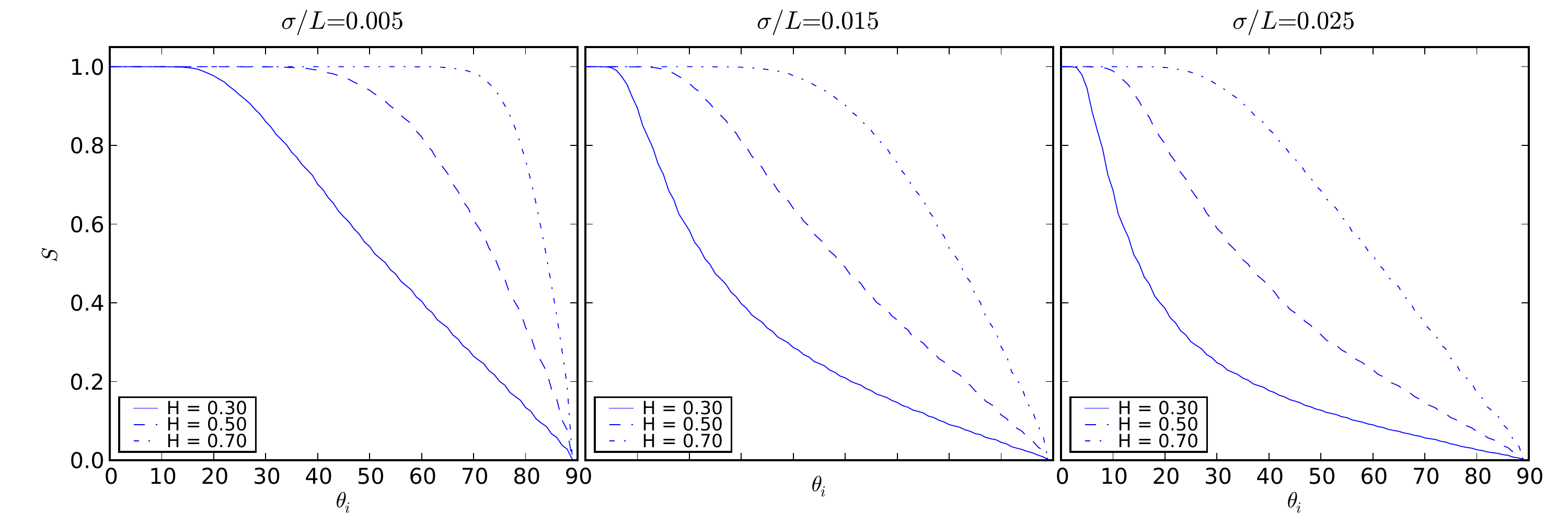}
	\caption{Nadir-viewed shadowing for fBm surfaces as a function of the 
		angle of incidence $\theta_i$.}
	\label{fig:shadowing_fbm_nadir}
\end{figure}

\begin{figure}
	\centering
	\includegraphics[width=0.85\textwidth]{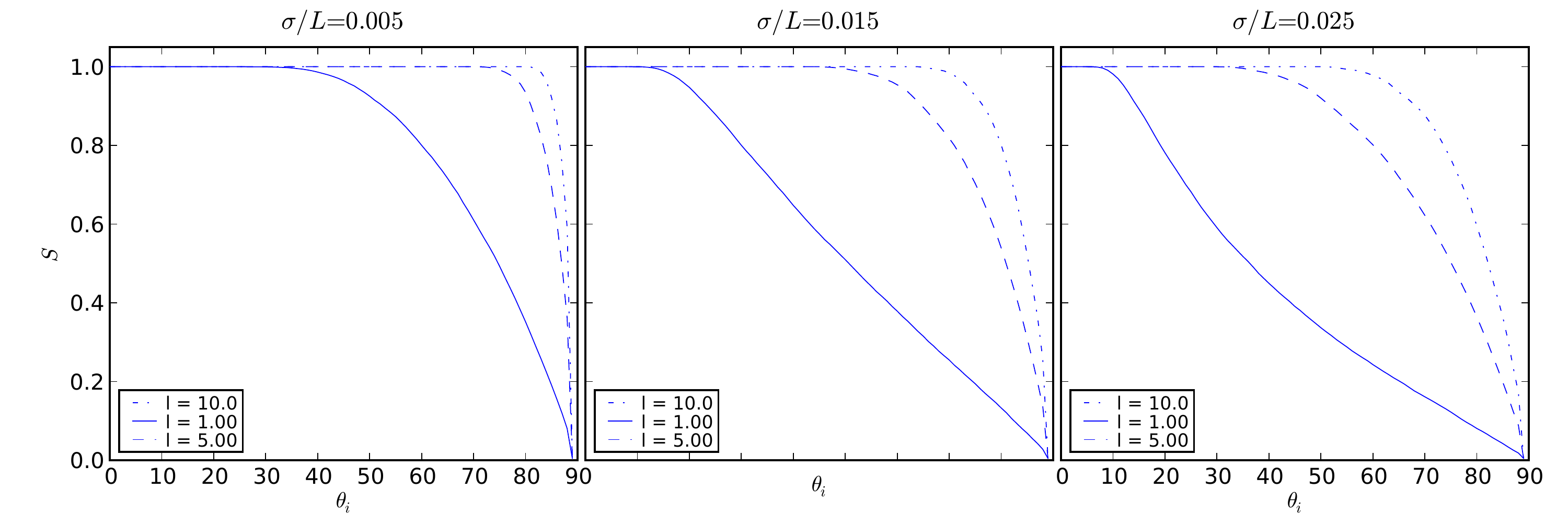}
	\caption{As in Fig. \ref{fig:shadowing_fbm_nadir} for surfaces with Gaussian correlation.}
	\label{fig:shadowing_gaussian_nadir}
\end{figure}

\begin{figure}
	\centering
	\includegraphics[width=.85\textwidth]{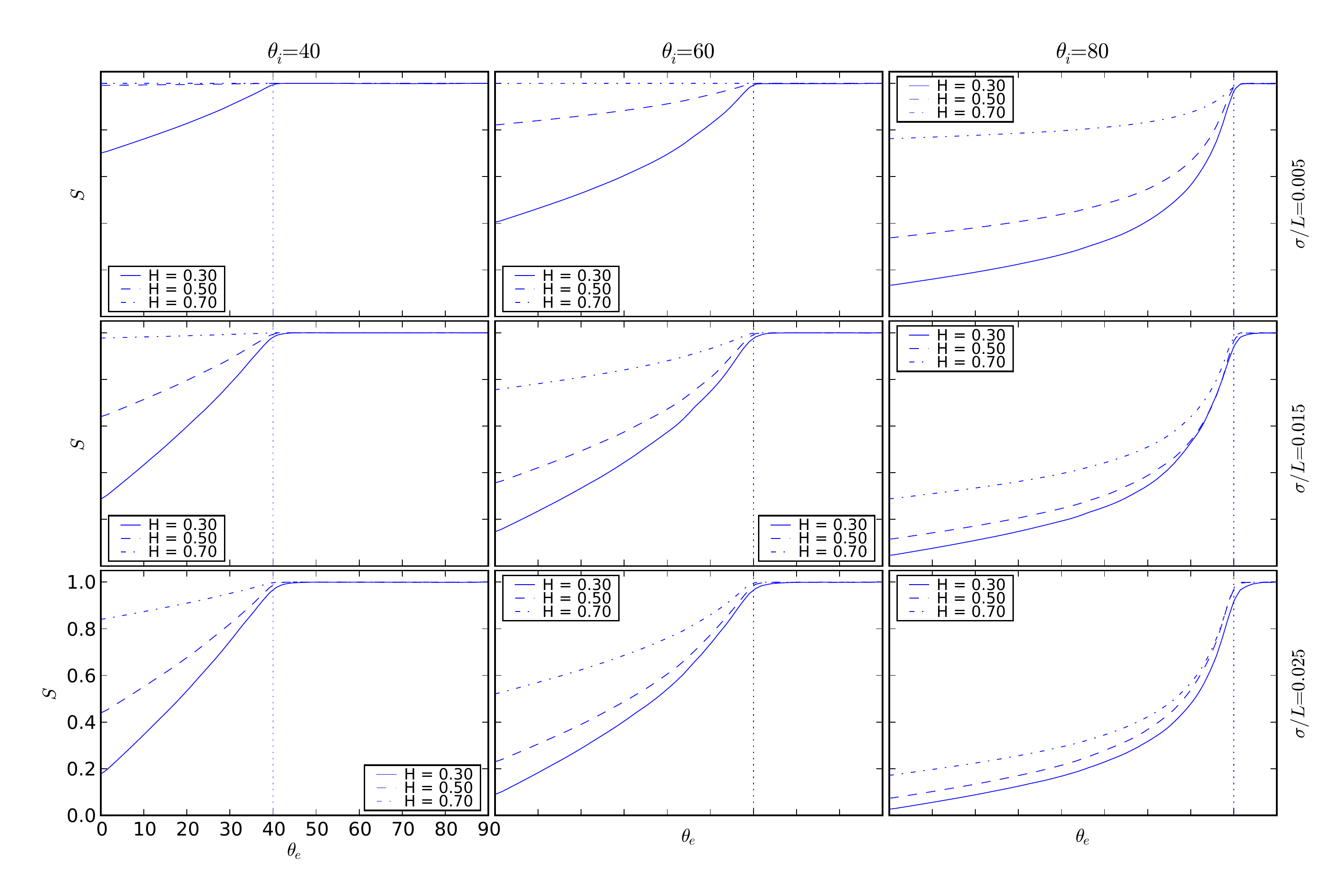}
	\caption{Combined shadowing and masking function along $\phi_e = 0$ for fBm surfaces as a function
of the 
		angle of emergent radiation $\theta_e$. Shown are plots for three different
		angles of incident radiation $\theta_i$, three different $\frac{\sigma}{L}$, and three 
		different $H$. The backscattering angle is drawn for each plot as a dotted
		vertical line.}
	\label{fig:shadowing_fbm_p00}
\end{figure}

\begin{figure}
	\centering
	\includegraphics[width=.85\textwidth]{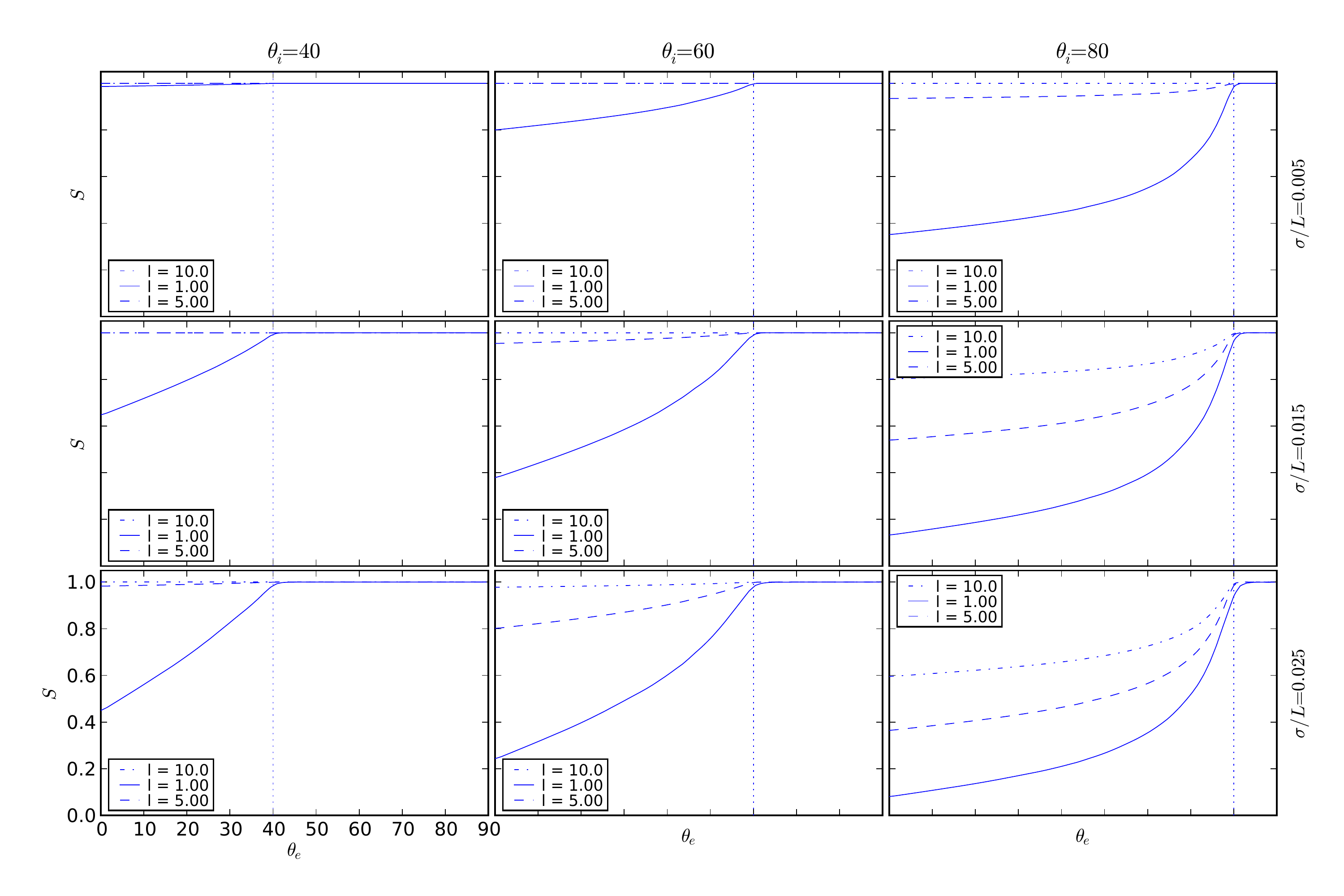}
	\caption{As in Fig. \ref{fig:shadowing_fbm_p00} for surfaces with Gaussian correlation.}
	\label{fig:shadowing_gaussian_p00}
\end{figure}

\begin{figure}
	\centering
	\includegraphics[width=.85\textwidth]{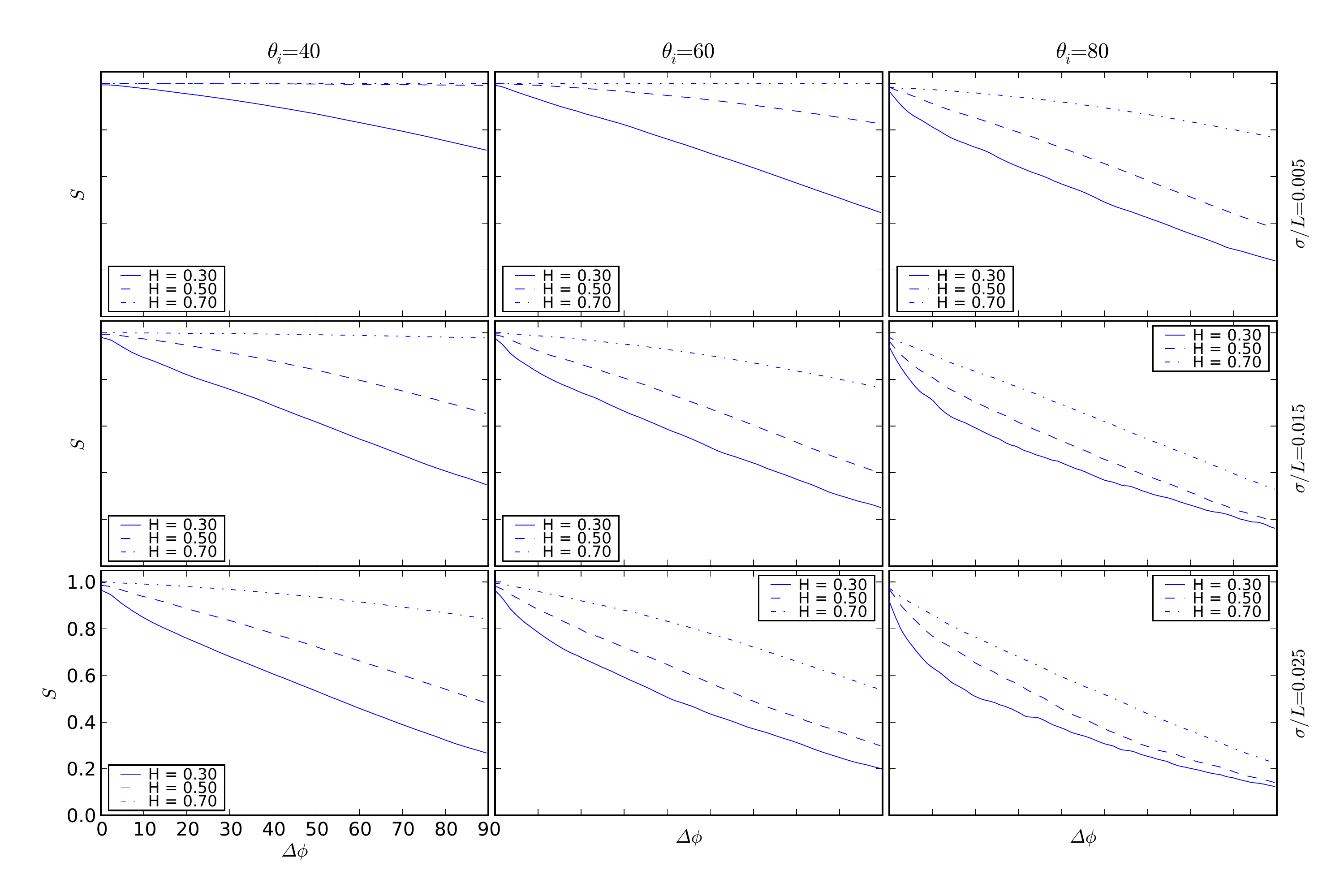}
	\caption{Shadowing for fBm-surfaces
		as a function of azimuth angle $\phi_e$ and constant $\theta_e = \theta_i$. The shape of 
		the azimuthal shadowing effect is seen as a function of varying surface 
		geometry. }
	\label{fig:shadowing_fbm_phi}
\end{figure}

\begin{figure}
	\centering
	\includegraphics[width=.85\textwidth]{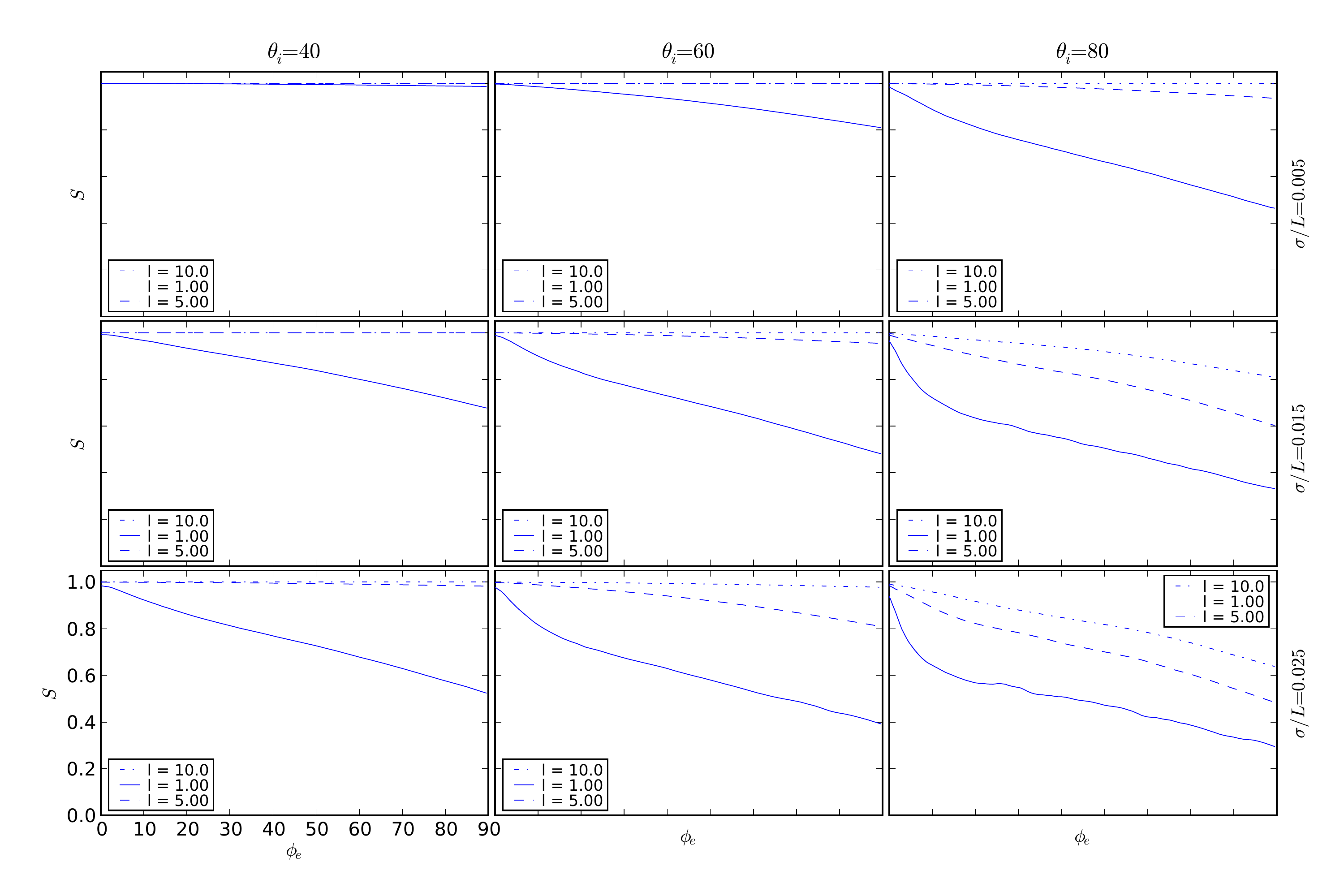}
	\caption{As in Fig. \ref{fig:shadowing_fbm_phi} for surfaces with Gaussian correlation.}
	\label{fig:shadowing_gaussian_phi}
\end{figure}

The nadir-viewed geometric shadowing shown in Figs. \ref{fig:shadowing_fbm_nadir} and
\ref{fig:shadowing_gaussian_nadir} can be compared to the study made by Shepard and
Campbell \cite{Shepard98}. The main result visible from the figures is well known: the steepness
of the shadowing function depends strongly on the scale of the surface roughness features.
Small-scale roughness has a greater contribution to the shadowing than the large-scale roughness
of equivalent amplitude.
Figures \ref{fig:shadowing_fbm_p00} and \ref{fig:shadowing_gaussian_p00} show the shadowing function
along constant azimuth angle $\phi_e = 0$, as a function of the angle of emergent radiation
$\theta_e$. As $\theta_e$ approaches the backscattering direction $\theta_i$, the shadowing function
approaches unity. For $\theta_e > \theta_i$, no shadows are visible. Small-scale roughness
($H=0.3$ or $\frac{l}{L} = 0.01$) can be seen to have a notable effect for even modest roughness
amplitudes. For the most extreme cases where $\frac{\sigma}{L} = 0.025$, shadowing in all the
simulated fBm surfaces approaches a similar form together with increasing $\theta_i$.
Figures \ref{fig:shadowing_fbm_phi} and \ref{fig:shadowing_gaussian_phi} show the shape of the
azimuthal shadowing effect along constant $\theta_e = \theta_i$ as a function of the azimuth angle
$\phi_e$. For most situations, the shadowing function can be considered linear for $ 20^\circ <
\Delta \phi < 90^\circ$.

\subsection{Lambertian Surfaces}

Next, the behaviour of rough surfaces with Lambertian scattering elements was considered. The
simulations were carried out using the same sets of parameters as for the pure shadowing-function
studies. Now, the observed radiance is a function of the distribution of the normals of the visible
unshadowed surface elements, as well as of the ratio of visible unshadowed surface elements to
the total visible surface area.

\begin{figure}
	\centering
	\includegraphics[width=0.8\textwidth]{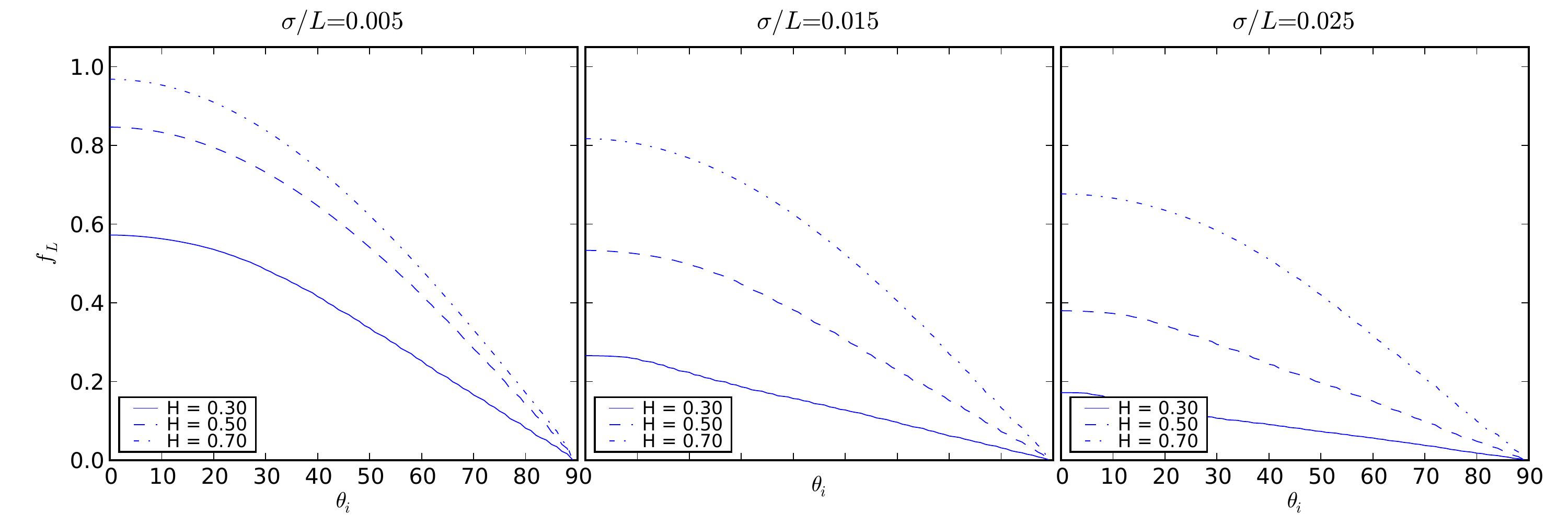}
	\caption{As in Fig. \ref{fig:shadowing_fbm_nadir} for a surface with Lambertian scattering
		elements.}
	\label{fig:fbm_nadir_lambert}
\end{figure}

\begin{figure}
	\centering
	\includegraphics[width=0.8\textwidth]{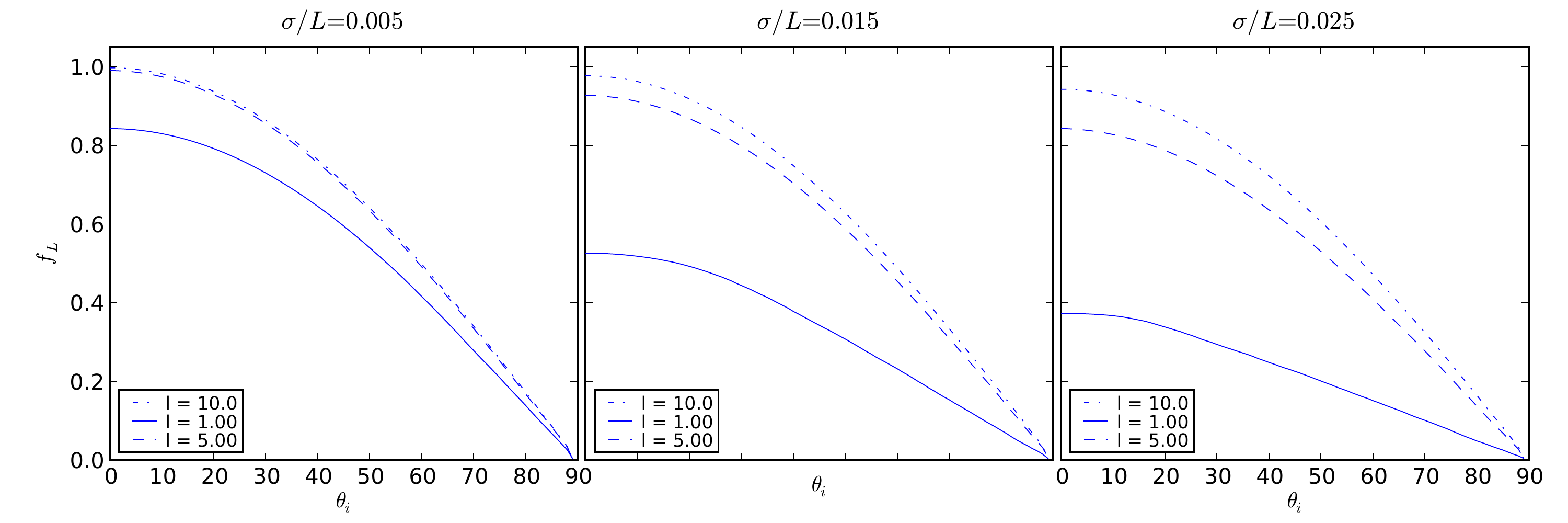}
	\caption{As in Fig. \ref{fig:shadowing_gaussian_nadir} for a surface with Lambertian scattering
		elements.}
	\label{fig:gaussian_nadir_lambert}
\end{figure}

\begin{figure}
	\centering
	\includegraphics[width=.85\textwidth]{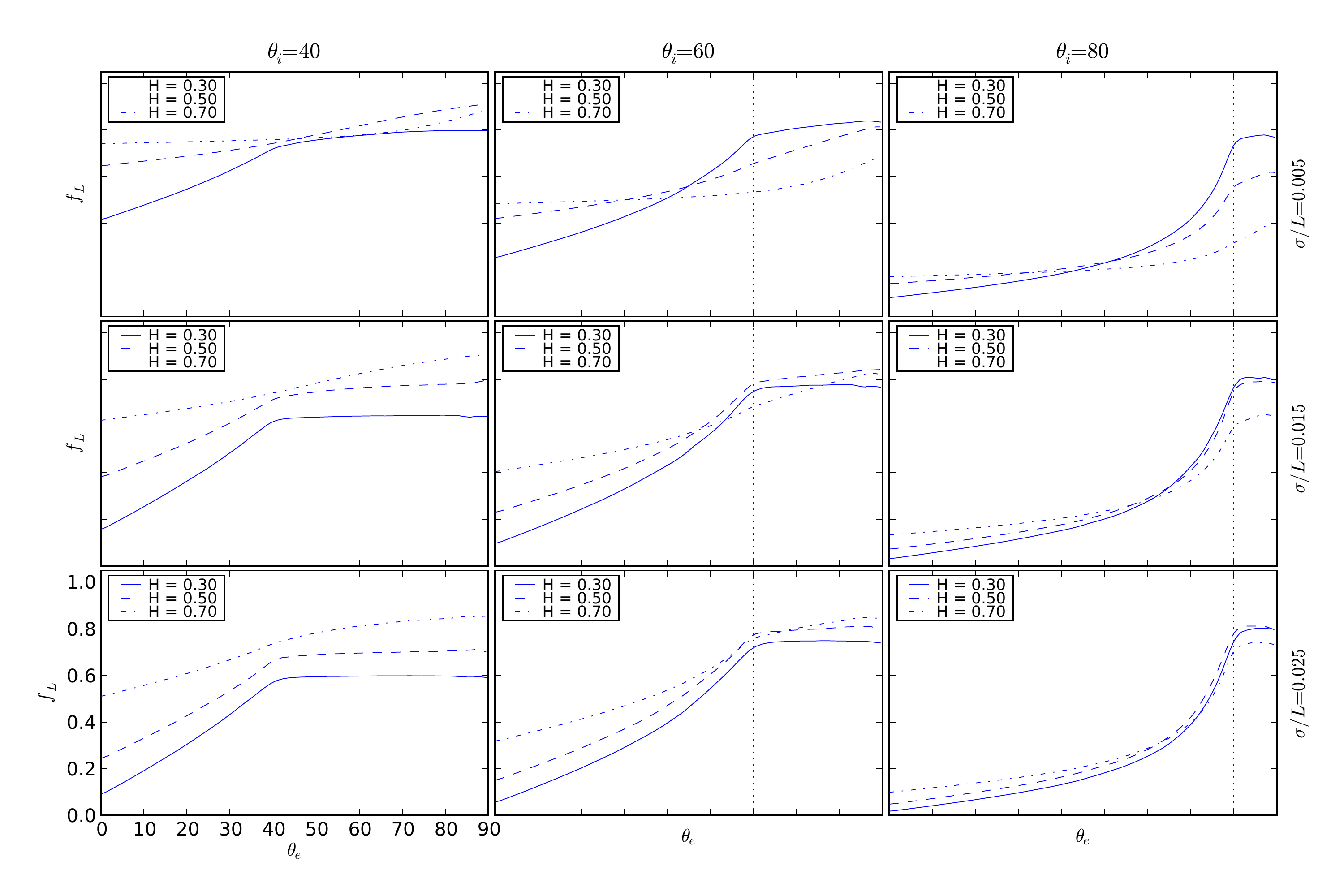}
	\caption{As in Fig. \ref{fig:shadowing_fbm_p00} for a surface with Lambertian scattering
		elements.}
	\label{fig:lambert_fbm_p00}
\end{figure}

\begin{figure}
	\centering
	\includegraphics[width=.85\textwidth]{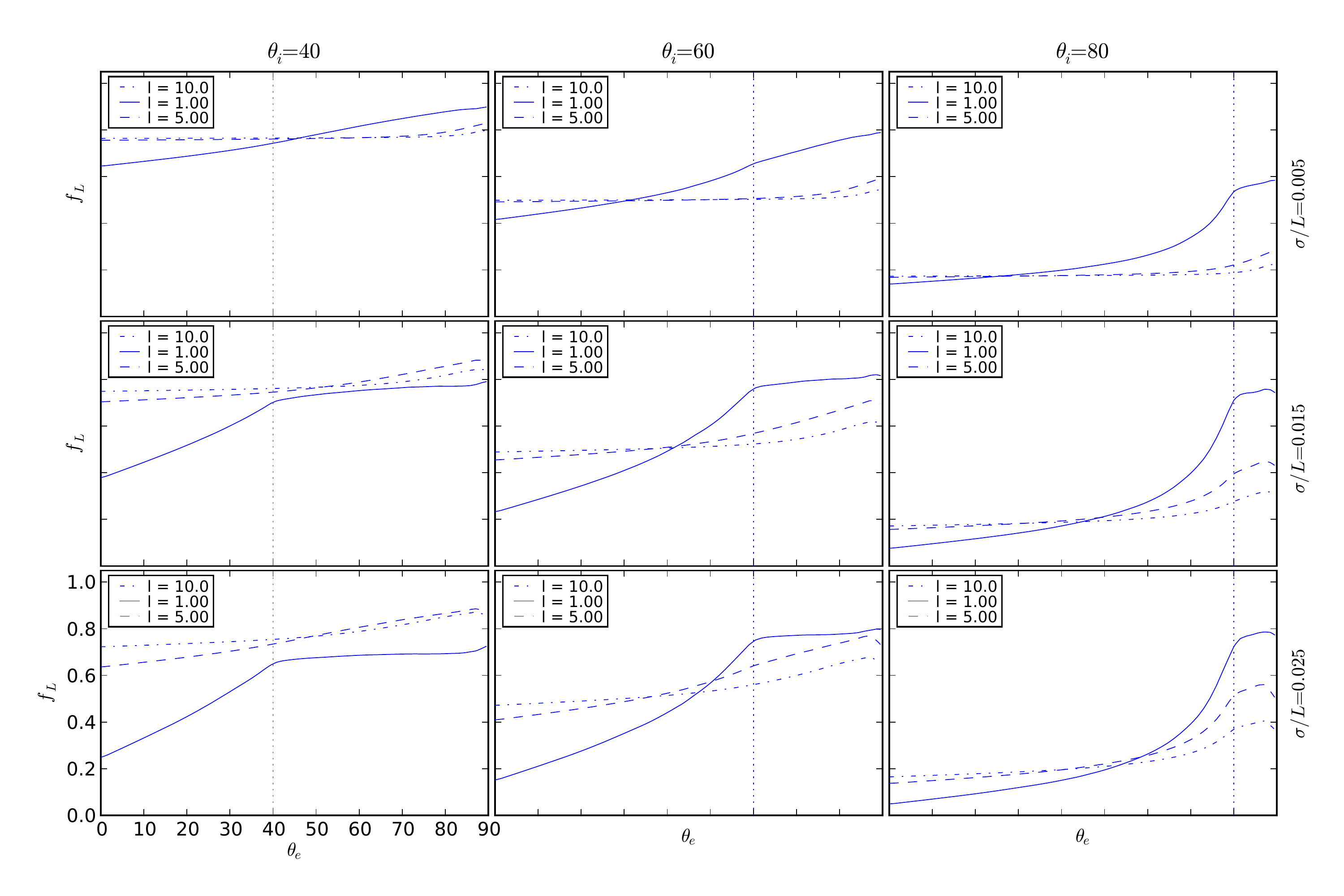}
	\caption{As in Fig. \ref{fig:shadowing_gaussian_p00} for a surface with Lambertian scattering
		elements..}
	\label{fig:lambert_gaussian_p00}
\end{figure}

\begin{figure}
	\centering
	\includegraphics[width=.85\textwidth]{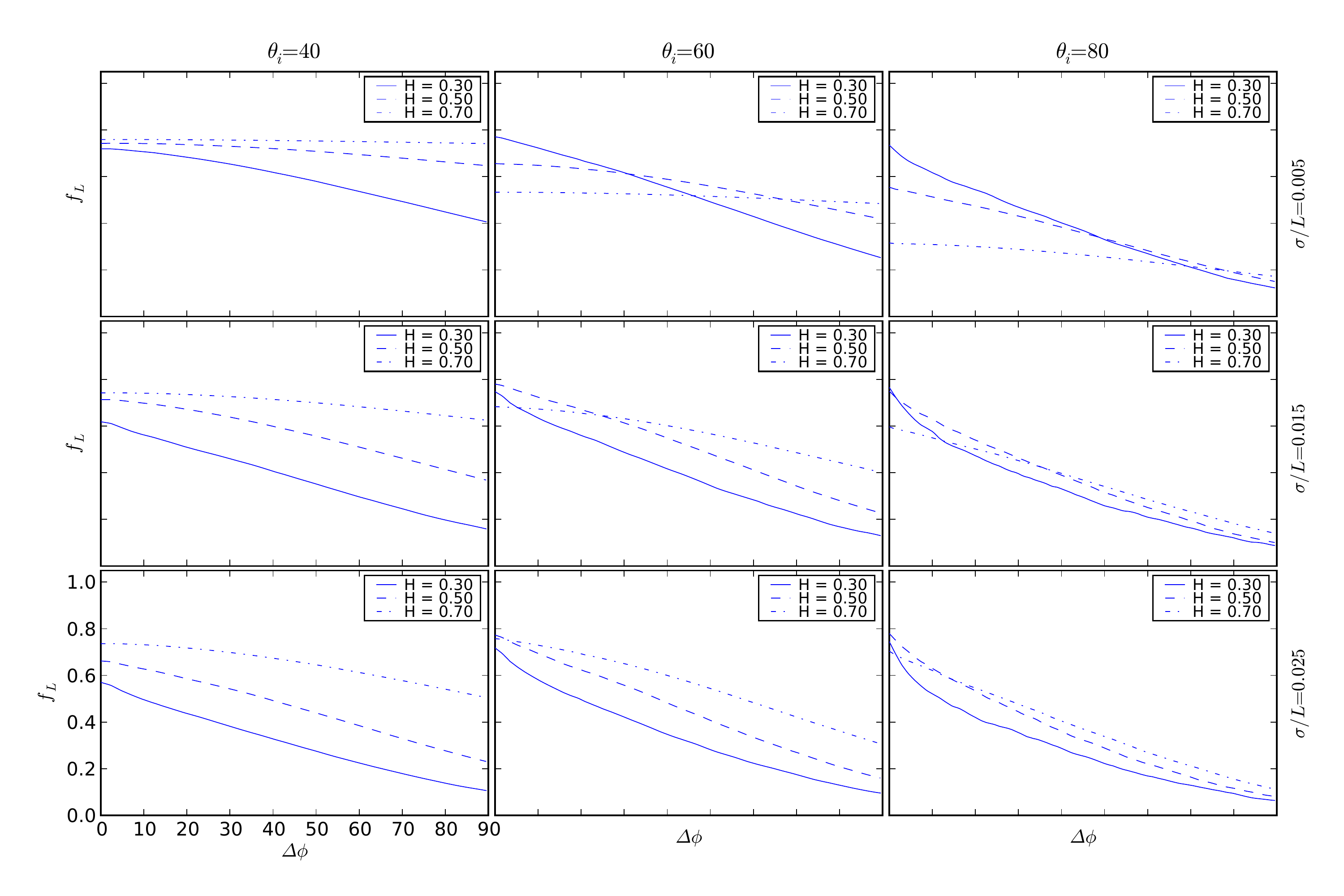}
	\caption{As in Fig. \ref{fig:shadowing_fbm_phi} for a surface with Lambertian scattering
		elements.}
	\label{fig:lambert_fbm_phi}
\end{figure}

\begin{figure}
	\centering
	\includegraphics[width=.85\textwidth]{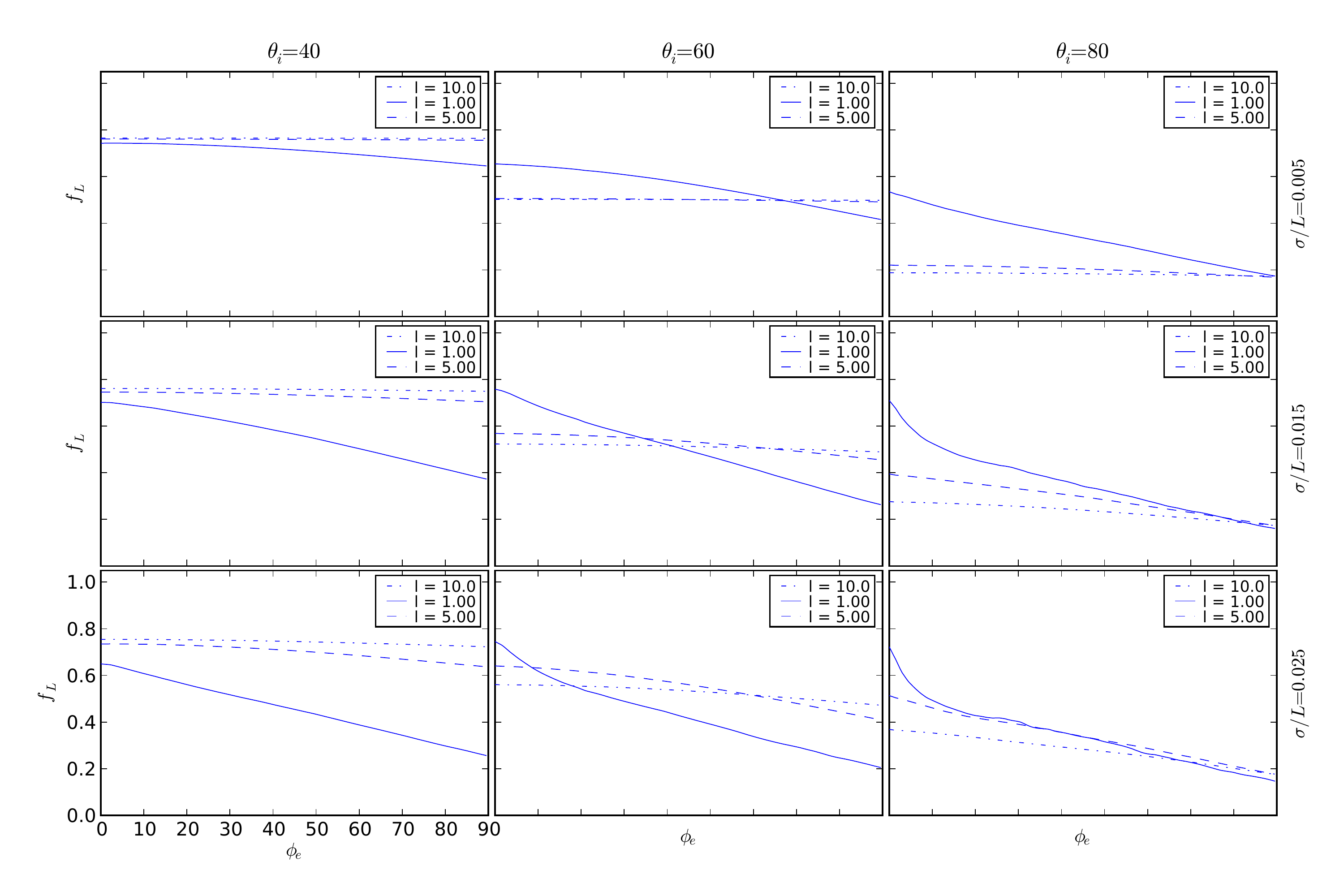}
	\caption{As in Fig. \ref{fig:shadowing_gaussian_phi} for a surface with Lambertian scattering
		elements.}
	\label{fig:lambert_gaussian_phi}
\end{figure}

The nadir-viewed rough-surface-corrected Lambertian BRDF $f_{L,RS}$ multiplied by $4\pi$ is shown in
Figs. \ref{fig:fbm_nadir_lambert} and \ref{fig:gaussian_nadir_lambert}. The $z$-component
of the normals of the surface elements decreases with increasing surface roughness amplitude, i.e.,
the rms-slope of the surface increases, and the radiance observed from the nadir decreases for all
$\theta_i$. For surfaces with Gaussian correlation function, only the smallest-scale roughness
$(\frac{l}{L} = 0.01)$ has a significant effect on the shape of the curve. For fBm surfaces, the
smaller values of $H$ yield more notable effects, as can be expected. The similarity between the
nadir-viewed values of fBm surfaces with $H=0.5$ and Gaussian correlation surfaces with $\frac{l}{L}
= 0.01$ is interesting.
Figures \ref{fig:lambert_fbm_p00} and \ref{fig:lambert_gaussian_p00} show $f_{L,RS}$ in a similar
fashion to Figs. \ref{fig:shadowing_fbm_p00} and \ref{fig:shadowing_gaussian_p00}. For a smooth
surface, the value of $f_L$ is constant for constant $\theta_i$. This is also the case for surfaces
with only large-scale roughness features of small amplitude. Small-scale roughness creates rather a
sharp change in the slope of the curve in the backscattering direction $\theta_e = \theta_i$. 
Figures \ref{fig:lambert_fbm_phi} and \ref{fig:lambert_gaussian_phi} show the behaviour of the
$f_{L,RS}$ as a function of the azimuth-angle. The azimuthal behaviour is still similar for the
fBm surfaces with $H=0.5$ and Gaussian correlation surfaces with  $\frac{l}{L}= 0.01$, but breaks
down when moving to large $\theta_i$ and extreme roughness amplitudes.

\subsection{Lommel-Seeliger Surfaces}

Finally, the behaviour of rough surfaces with Lommel-Seeliger scattering elements was studied.
As for the Lambertian BRDF, the simulations were carried out using the same sets of parameters as
for the pure shadowing function studies.

\begin{figure}
	\centering
	\includegraphics[width=0.8\textwidth]{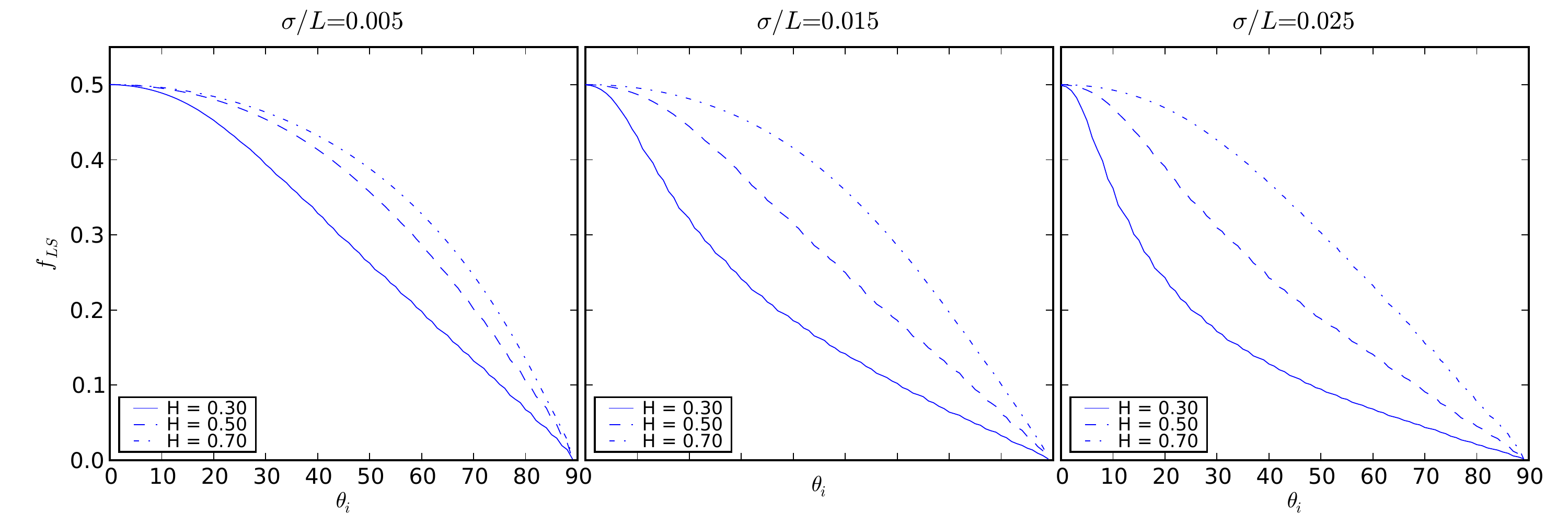}
	\caption{As in Fig. \ref{fig:shadowing_fbm_nadir} for a surface with Lommel-Seeliger scattering
		elements.}
	\label{fig:fbm_nadir_lommelseeliger}
\end{figure}

\begin{figure}
	\centering
	\includegraphics[width=0.8\textwidth]{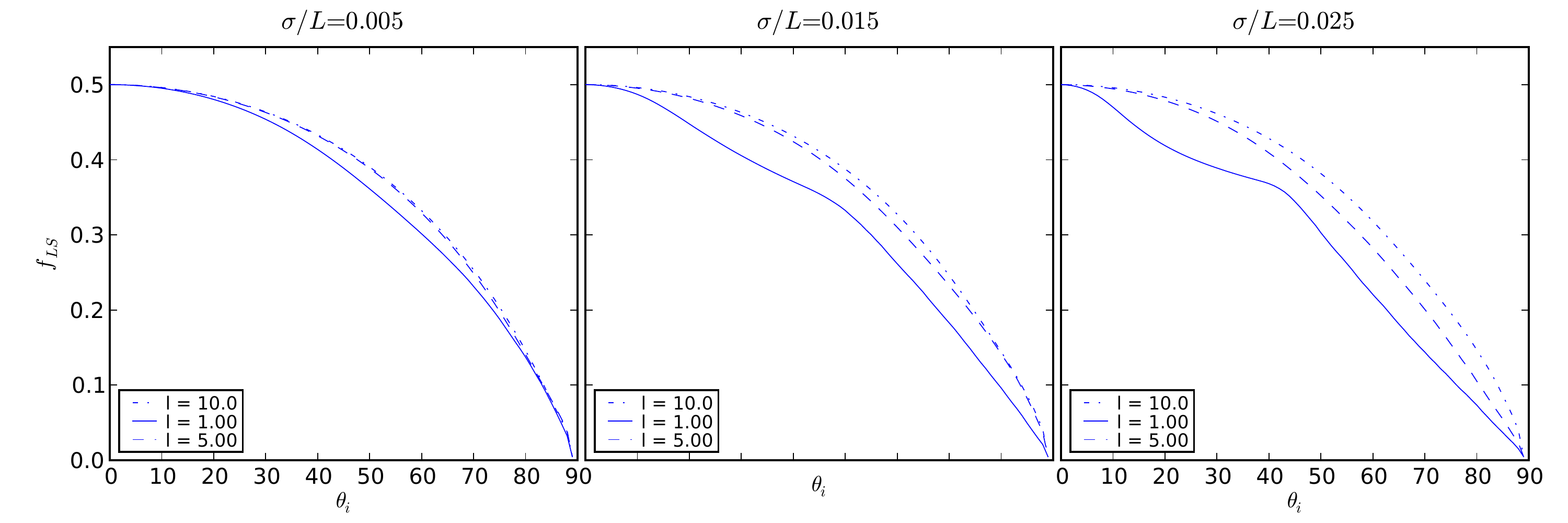}
	\caption{As in Fig. \ref{fig:shadowing_gaussian_nadir} for a surface with Lommel-Seeliger
		scattering elements.}
	\label{fig:gaussian_nadir_lommelseeliger}
\end{figure}

\begin{figure}
	\centering
	\includegraphics[width=.85\textwidth]{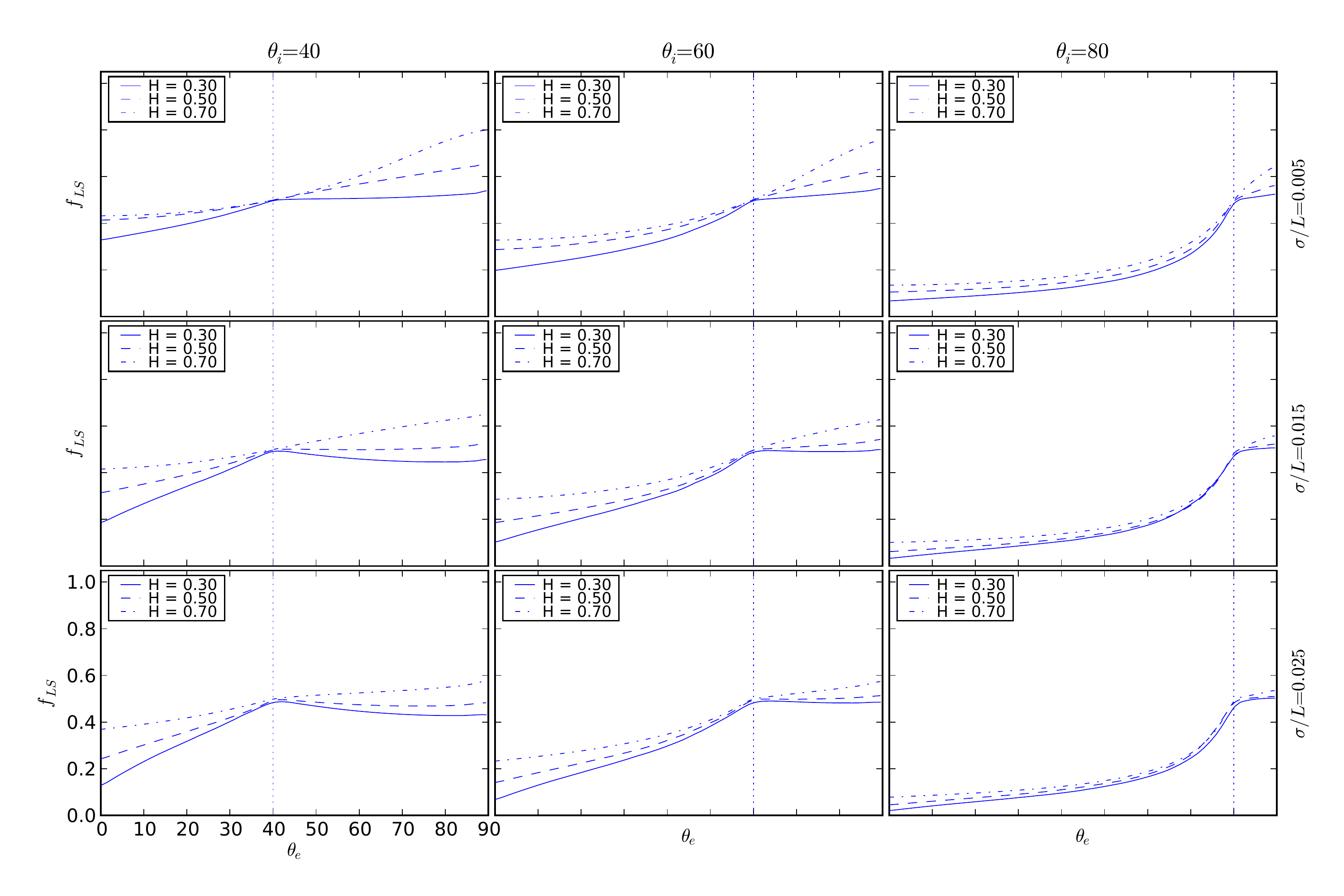}
	\caption{As in Fig. \ref{fig:shadowing_fbm_p00} for a surface with Lommel-Seeliger scattering
		elements.}
	\label{fig:lommelseeliger_fbm_p00}
\end{figure}

\begin{figure}
	\centering
	\includegraphics[width=.85\textwidth]{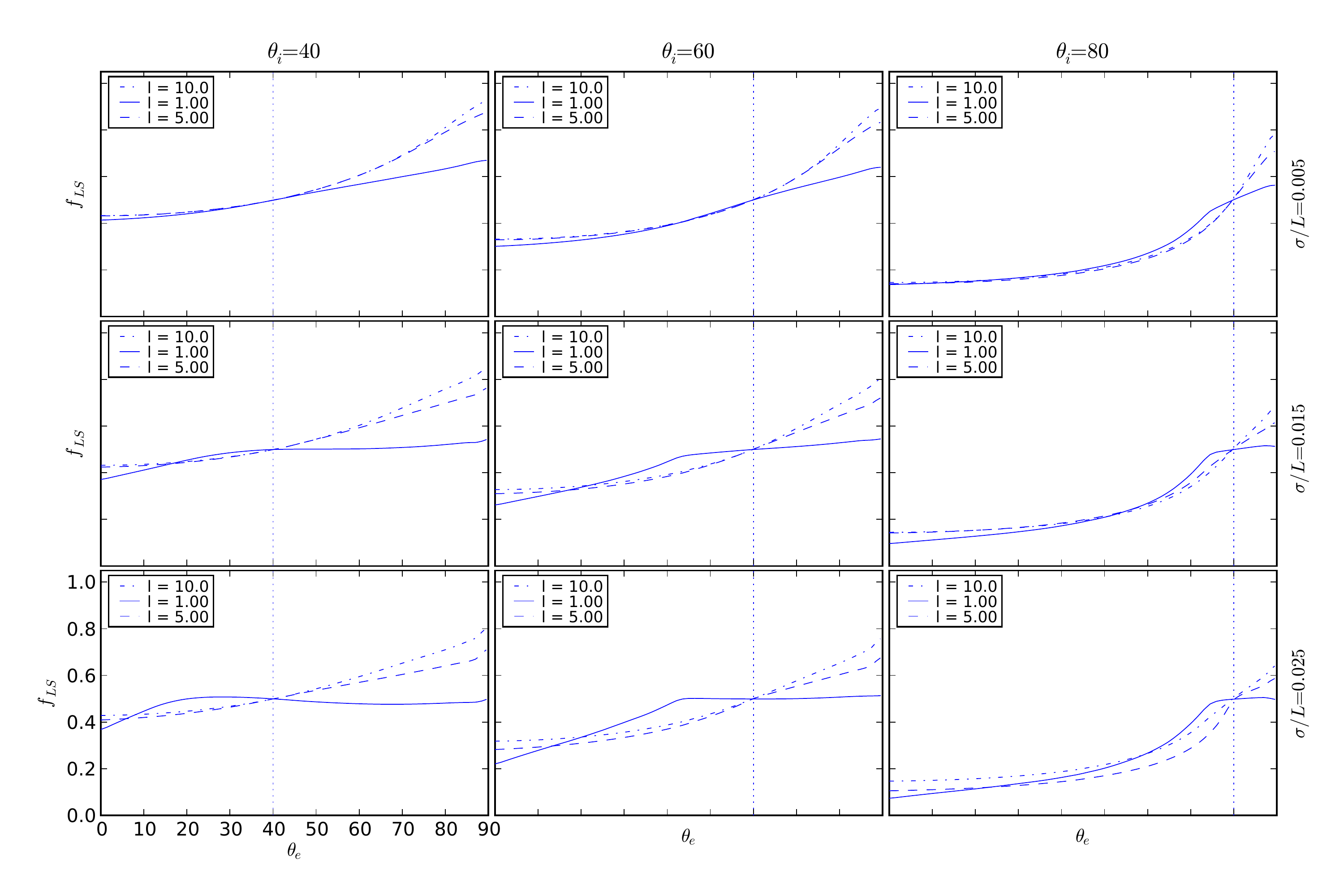}
	\caption{As in Fig. \ref{fig:shadowing_gaussian_p00} for a surface with Lommel-Seeliger
scattering
		elements.}
	\label{fig:lommelseeliger_gaussian_p00}
\end{figure}

\begin{figure}
	\centering
	\includegraphics[width=.82\textwidth]{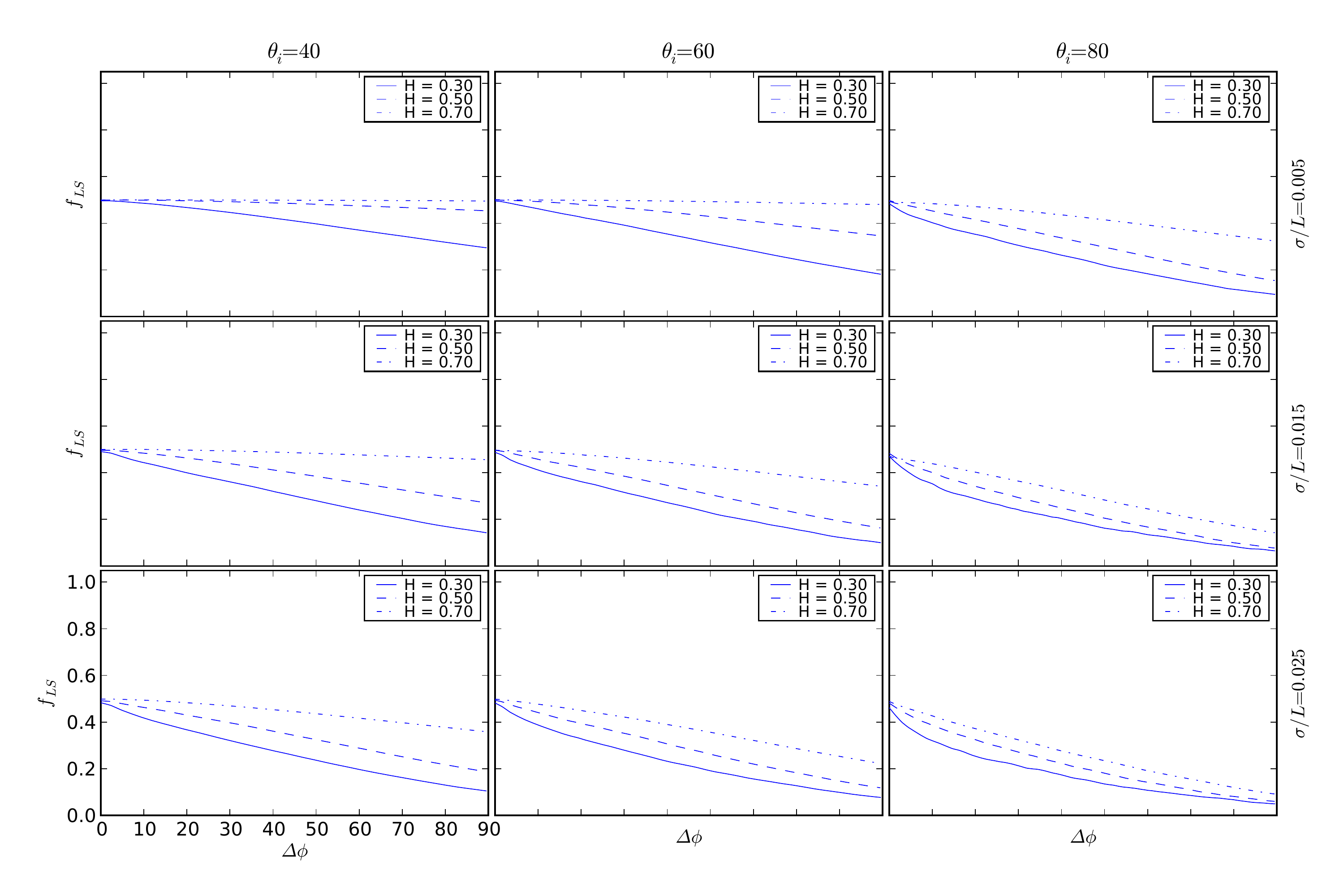}
	\caption{As in Fig. \ref{fig:shadowing_fbm_phi} for a surface with Lommel-Seeliger scattering
		elements.}
	\label{fig:lommelseeliger_fbm_phi}
\end{figure}

\begin{figure}
	\centering
	\includegraphics[width=.82\textwidth]{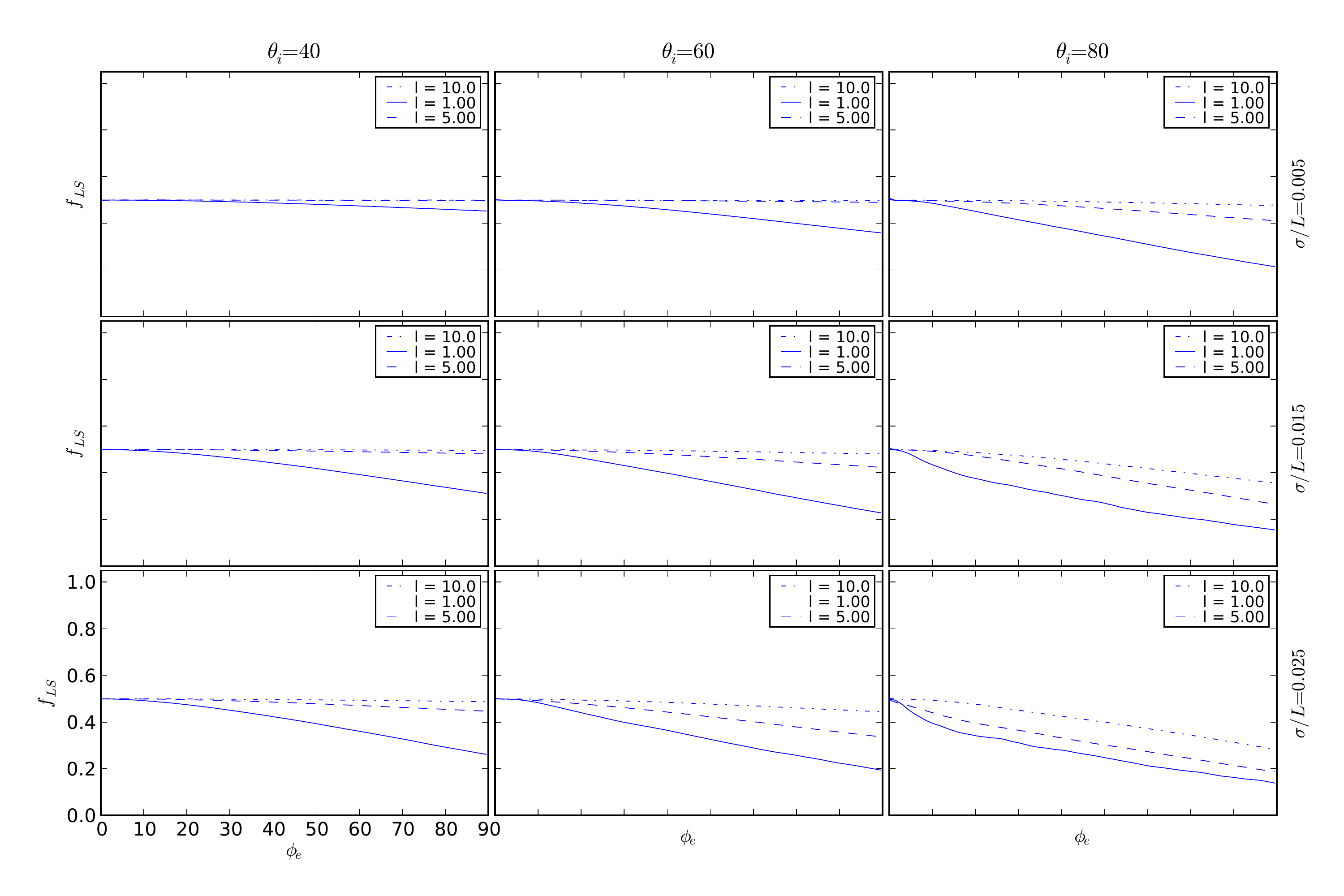}
	\caption{As in Fig. \ref{fig:shadowing_gaussian_phi} for a surface with Lommel-Seeliger
scattering
		elements.}
	\label{fig:lommelseeliger_gaussian_phi}
\end{figure}

The nadir-viewed rough-surface corrected Lommel-Seeliger BRDF $4\pi f_{LS,RS}$ is shown in Figs.
\ref{fig:fbm_nadir_lommelseeliger} and \ref{fig:gaussian_nadir_lommelseeliger}. The behaviour
differs both from the pure shadowing function and the rough-surface-corrected Lambertian BRDF. The
most notable feature is that for the backscattering geometry $\theta_e = \theta_i$, $4\pi f_{LS,RS}
= \frac{1}{2}$, something that can be also be seen from Figs. \ref{fig:lommelseeliger_fbm_p00},
\ref{fig:lommelseeliger_gaussian_p00}, \ref{fig:lommelseeliger_fbm_phi}, and
\ref{fig:lommelseeliger_gaussian_phi}. This is due to the $\frac{\mu_0}{\mu + \mu_0}$ dependence of
the Lommel-Seeliger scattering law. For the backscattering geometry with $\mu = \mu_0$, which
results in
$f_{LS,RS} = \frac{1}{4\pi}$, unlike for the Lambertian scattering model, the
distribution of surface slopes does not matter. 

\begin{figure}
	\centering
	\includegraphics[width=0.75\textwidth]{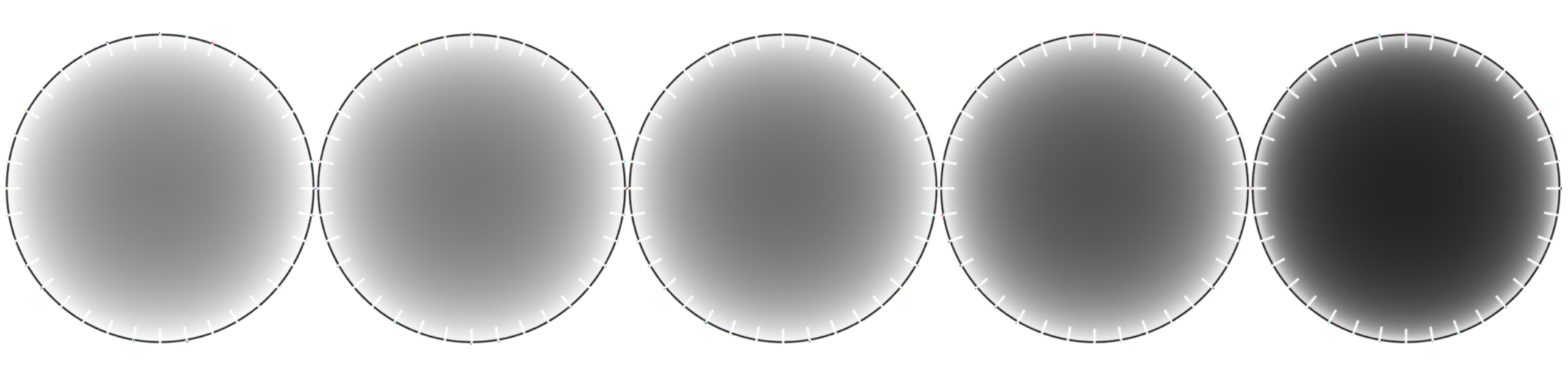}
	\caption{Lommel-Seeliger BRDF for 
	$\theta_i = 0^{\circ}, 20^{\circ}, 40^{\circ}, 60^{\circ}, 80^{\circ}$.}
	\label{fig:LS}
\end{figure}

\begin{figure}
	\centering
	\includegraphics[width=0.75\textwidth]{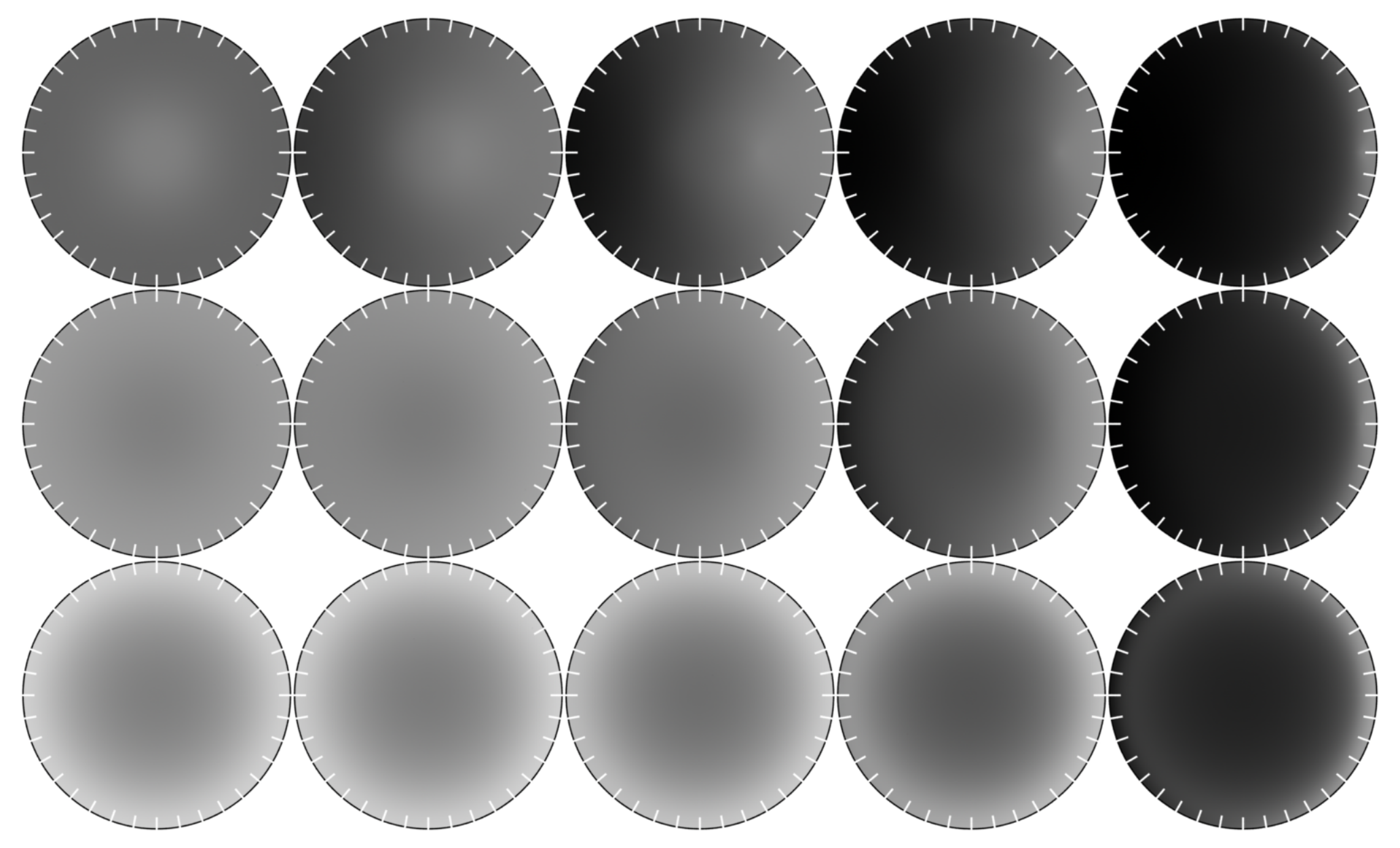}
	\caption{$f_{LS,RS}$ with $\frac{\sigma}{L} = 0.005$, $H = 0.3, 0.5, 0.7$, $\theta_i =
0^{\circ}, 20^{\circ}, 40^{\circ}, 60^{\circ}, 80^{\circ}$.}
	\label{fig:LS_RS_005}
\end{figure}

\begin{figure}
	\centering
	\includegraphics[width=0.75\textwidth]{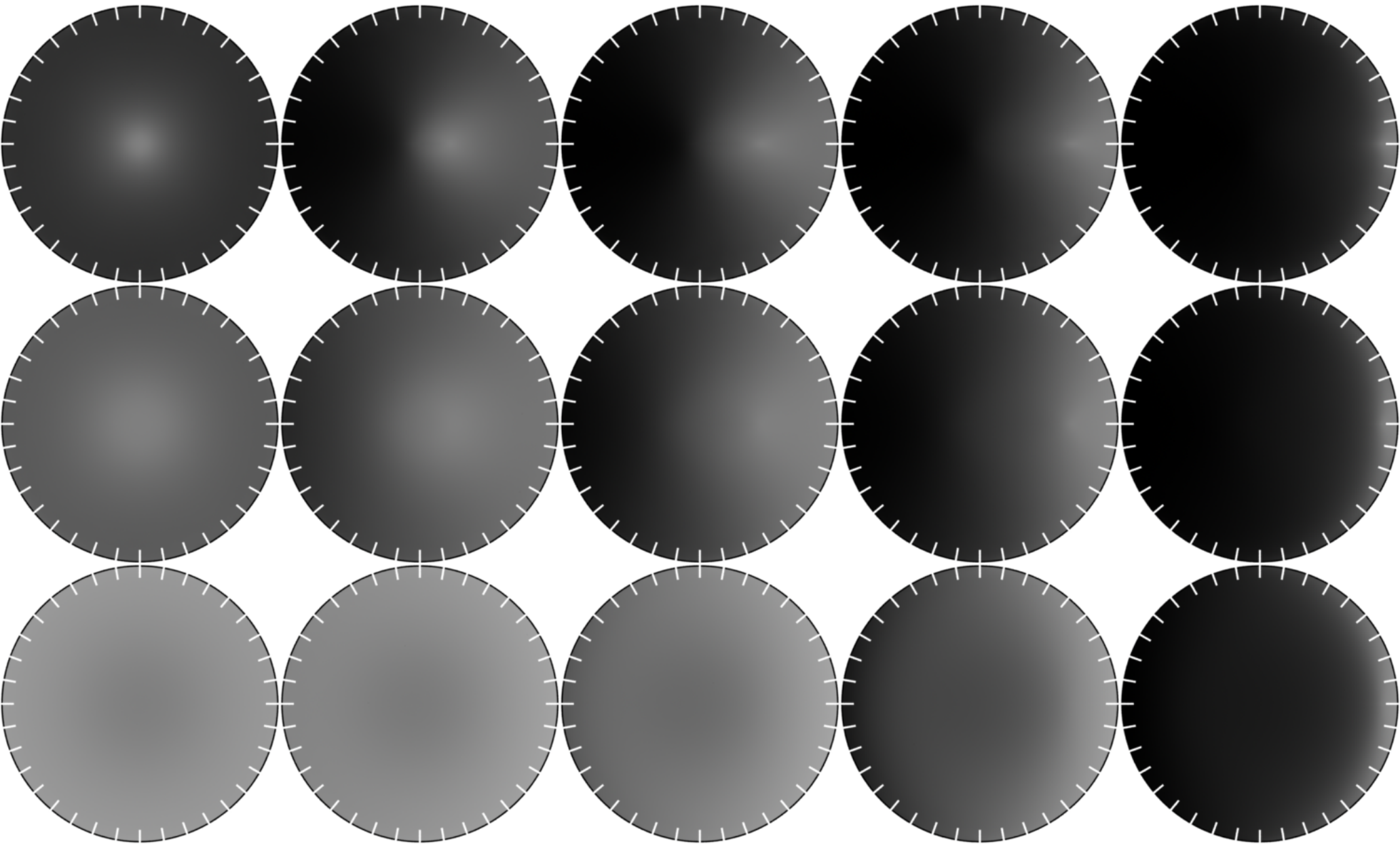}
	\caption{$f_{LS,RS}$ with $\frac{\sigma}{L} = 0.015$, $H = 0.3, 0.5, 0.7$, $\theta_i =
0^{\circ}, 20^{\circ}, 40^{\circ}, 60^{\circ}, 80^{\circ}$.}
	\label{fig:LS_RS_025}
\end{figure}

Figure \ref{fig:LS} shows the basic Lommel-Seeliger BRDF $4\pi f_{LS}$ mapped to spherical
coordinates $(\theta = \Delta \phi, r = \theta_e)$, $\Delta \phi = [0...2\pi]$, $\theta = [0...
\frac{\pi}{2}]$. The brightness value is linearly mapped to $[0...1]$, black $= 0$, white $= 1$.
Figures \ref{fig:LS_RS_005} and \ref{fig:LS_RS_025} show the behaviour of $f_{LS,RS}$ for fBm
surfaces with $H = 0.3, 0.5, 0.7$, $\frac{\sigma}{L} = 0.005, 0.025$, and $\theta_i = 0^{\circ},
20^{\circ}, 40^{\circ}, 60^{\circ}, 80^{\circ}$. The geometric rough-surface shadowing effect is
clearly visible for surfaces with $H=0.3$, and can be distinguished as well from the surface
$H=0.5$, $\frac{\sigma}{L} = 0.015$.

\subsection{Discussion}

Rough-surface shadowing has been studied, together with rough-surface-corrected Lambert and
Lommel-Seeliger reflectance functions. Since the paper discusses only the first-order scattering
effects in the geometric-optics regime, it applies to bodies with low single-scattering albedo
$\tilde{\omega}$. The importance of multiple scattering increases together with increasing
$\tilde{\omega}$ and, for simulations of bright objects, computation of several orders of
scattering is necessary. Multiple scattering between surface roughness elements is considered to
fade the geometric shadowing effects and smoothen the reflectance model \cite{Shkuratov04}.
Inclusion of multiple scattering to the simulation would make $\tilde{\omega}$  an extra parameter
to the resulting scattering model, though different levels of scattering could be possible to
separate from each other.
Another important factor, when the surface-roughness features are small considered to
the mean free path of the radiation inside the scattering medium, is the subsurface scattering.
Subsurface scattering would also smoothen the effects from shadowing and masking, but would
include the absolute scale dependency, as well as the dependency on the scattering parameters of the
medium.

Shadowing arising from fBm surfaces depends on the Hurst-exponent $H$. Decreasing $H$ leads
to increasing power for small-scale roughness features. The smaller-scale roughness features
have a stronger effect on the shadowing of the surface. Thus, a rough surface with small $H$ and
relatively small surface-roughness amplitude shows much greater shadowing than a surface with larger
$H$ and roughness amplitude. For surfaces with only one major roughness scale, the angular
dependence of shadowing, when moving away from the opposition geometry, depends strongly on the
ratio $\frac{\sigma}{l}$. The behavior of shadowing from the opposition direction as a function of
the azimuth angle is almost linear in the range $\Delta \phi =[0...\frac{\pi}{2}]$ for fBm surfaces
with large $H$. For smaller $H$, a nonlinear brightening can be observed. The same results apply to
surfaces with Lommel-Seeliger scattering elements.

\section*{Acknowledgements}
We thank the anonymous referee for helpful and costructive comments. Research was supported,
in part, by the Academy of Finland.

\bibliographystyle{elsart-num}
\bibliography{Rough_Surface_Scattering}

\end{document}